\documentclass[12pt]{iopart}

\usepackage{graphicx}
\usepackage{cite}
\usepackage{color}
\usepackage{multirow}

\begin{document}

\setlength{\parindent}{0pt}

\title[The effect of photoemission nsec He microdischarges]{The effect of photoemission on nanosecond helium microdischarges at atmospheric pressure}

\author{Zolt\'an Donk\'o$^{1,2}$, Satoshi Hamaguchi$^2$, Timo Gans$^3$}

\address{$^1$Institute for Solid State Physics and Optics, Wigner Research Centre for Physics, Hungarian Academy of Sciences, 1121 Budapest, Konkoly Thege Mikl\'os str. 29-33, Hungary\\
$^2$Center for Atomic and Molecular Technologies,
Graduate School of Engineering, Osaka University,
2-1 Yamadaoka, Suita, Osaka 565-0871, Japan\\
$^3$York Plasma Institute, Department of Physics, University of York, Heslington, York, United Kingdom}
\ead{donko.zoltan@wigner.mta.hu}

\begin{abstract}
Atmospheric-pressure microdischarges excited by nanosecond high-voltage pulses are investigated in helium-nitrogen mixtures by first-principles particle-based simulations that include VUV resonance radiation transport via tracing photon trajectories. The VUV photons, of which the frequency redistribution in emission processes is included in some detail, are found to modify remarkably the computed discharge characteristics due to their ability to induce electron emission from the cathode surface. The electrons created this way enhance the plasma density and a significant increase of the transient current pulse amplitude is observed. The simulations allow the computation of the density of helium atoms in the 2$^1$P resonant state, as well as the density of photons in the plasma and the line shape of the resonant VUV radiation reaching the electrodes. These indicate the presence of significant radiation trapping in the plasma and photon escape times longer than the duration of the excitation pulses are found.
\end{abstract}


\submitto{\PSST}
\maketitle

\section{Introduction}

Short-pulse discharges excited with $\sim$ kV voltages in various gases have been attracting considerable interest. Such systems have been studied in a wide pressure range extending from low ($\sim$ mbar) to atmospheric pressures \cite{Svetlana}. The width of the excitation voltage pulses ranges from the picosecond domain up to tens, or hundreds of nanoseconds \cite{Svetlana2,Starikovskiy}. The gases used include pure noble gases and their mixtures with molecular gases, as well as air, which is particularly important for various practical applications. 

Advanced diagnostics tools, like Thomson scattering and laser absorption spectroscopy measurements allow understanding the ignition processes and the dynamics of the afterglow in high pressure nanosecond pulsed helium micro-discharges \cite{Uwe}. Recent studies addressed the effect of surface protrusion on plasma sheath properties \cite{Fu} and the effect of the pulse rise time on the discharge characteristics, especially the breakdown voltage \cite{Zhang}.

Besides their rich physics, short-pulse discharges have various applications, e.g., their switching applications have been explored in \cite{Bokhan,Bokhan2} and the effects of their plasma on cells and tissues have been discussed in \cite{Beebe}. 

At high pressures, computations based on hydrodynamic models are much more efficient than kinetic simulations, due to the very high collisionality of the plasma, allowing detailed considerations of the chemical kinetics \cite{Waskoenig,Niemi}. Nonetheless, particle-based kinetic approaches, like the Particle-in-Cell method complemented with Monte Carlo Collisions (PIC/MCC), have also been applied in some studies, despite the high computational requirements, because they can describe non-local kinetic effects in domains with high reduced electric field and they provide access to the electron energy distribution function. Such studies include, e.g., those of fast (subnanosecond) breakdown in high-voltage open discharges \cite{Schweigert,Schweigert2} and of the discharge development in hydrogen micordischarges \cite{Donko-h2,CW1,CW2}, as well as ionisation instabilities \cite{Lieberman} and self-organised pattern formation \cite{Weili}. 

The ultraviolet (UV) / vacuum-ultraviolet (VUV) radiation of the plasma plays an important role in various applications. UV/VUV photons in plasma technologies have various effects, thus controlling their fluxes (via setting operation parameters and gas mixing ratios) is of great importance \cite{Tian}. Plasma display panels \cite{Boeuf} operate on the basis of the generation of VUV radiation in medium-pressure noble gas mixture (dielectric barrier) discharges that is, in turn, converted to visible light by different phosphors designed for different wavelength domains. The UV/VUV radiation of dielectric barrier discharges of different types may assist decomposing various hazardous organic molecules \cite{Kogelschatz}, while similarly short-wavelength radiation from high-frequency (primarily microwave) discharges and their afterglows has widely been used for sterilization \cite{Sterilization}. The effect of VUV radiation of plasma jets on reactive oxygen species generation in bio-relevant liquids was investigated in \cite{Jablonowski}, furthermore, the ultraviolet radiation of plasmas was confirmed to play an important role in the creation of excited hydroxyl radicals under the conditions applicable to plasma-based cancer treatment \cite{cancer}. 

All the applications mentioned above call for an accurate description of UV/VUV photon transport in discharge plasmas. In high-pressure transient plasmas this transport is "strongly coupled" to the (charged particle) dynamics of the discharges as the creation, transport and arrival of the photons at the electrodes can have an influence on the discharge, which in turn will modify the spatio-temporal creation rates of excited states and consequent emission of photons. Moreover, under such conditions, the characteristic escape time of the VUV photons from the plasma may be comparable to, or longer than the time scales of the discharge pulses. Thus, de-coupling of the modelling of photon transport from the modelling of the transport of charged particles should be avoided at such conditions.

In this work, we report a simulation study, based on the first-principle PIC/MCC approach, that also includes a particle-based treatment of VUV resonance photons, of atmospheric pressure microdischarges created in He gas with a small admixtures of N$_2$. The discharge model and its implementation is described in section 2. Section 3 presents our results, in section 3.1 the physics of the discharge is discussed in details for a given set of conditions and parameters, while in section 3.2 some results of a parameter variation are presented. Section 4 gives a summary of the work.

\section{Simulation model and method}

We simulate the formation of atmospheric-pressure microplasmas by the PIC/MCC technique. As an important and significant addition, particle treatment of VUV photons is incorporated into the code. Concerning the latter, we employ the approach of Fierro {\it et al.} \cite{Fierro}, who studied the effect of photoemission induced by the resonance radiation of the plasma on the electron current in self-sustained He dc discharges. The "full" particle treatment of the transport of both charged particles and photons is computationally very intensive, but its unique advantage that it is based on the first principles makes it an important alternative to (more efficient, but more approximate) fluid approaches. Below, we give the details of the discharge model and the simulation method. The treatment of the charged and excited species is outlined in section 2.1, the description of the VUV radiation transport is discussed in section 2.2, while timing issues and general features of the simulations are overviewed in section 2.3.

\subsection{Charged species and reaction processes}

Our PIC/MCC code is one-dimensional in real space and considers the axial and radial components of the velocity. The charged particles considered in the code are electrons, He$^+$ ions, He$_2^+$ ions, and N$_2^+$ ions. Time is discretised in the simulation and between consecutive collisions the particles move along trajectories that result from the solution of their equations of motion. Following a $\Delta t$  discrete time step used for the given projectile, the collision probability is found according to
\begin{equation}
P = 1- \exp(-\nu_{\rm proj, tot} \Delta t) = 1- \exp(- n_{\rm targ}\, \sigma_{\rm proj, tot}\, g \,\Delta t),
\label{eq:pcoll1}
\end{equation}
where $\nu_{\rm proj, tot}$ is the total collision frequency of the given projectile, $n_{\rm targ}$ is the density of the target species, $\sigma_{\rm proj, tot}$ is the total cross section of the interaction between the projectile and target species, $g$ is the relative velocity of the collision partners. In the case of ionic projectile species we take into account the thermal motion of the background gas particles via choosing random partners (from the Maxwell-Boltzmann distribution of the background atoms) in possible collision events. For electrons, we use the cold gas approximation, i.e., take $g$ to be equal to the projectile (electron) velocity. In order to optimize the runtime of the simulations we use different time steps for the different species (see section 2.3). The probability given by eq. (\ref{eq:pcoll1}) is compared to a random number $R_{01}$ that has a uniform distribution over the $[0,1)$ interval, and a collision is executed if $R_{01}<P$ is fulfilled. The type of collision, as well as the scattering and azimuth angles are also chosen based on the standard stochastic approach \cite{Donko2011}. The list of gas phase elementary processes, which are considered in the model and are discussed below, is listed in table 1. 

\begin{table}
\caption{\label{o2} List of gas phase elementary processes considered in the simulation.}
\footnotesize
\begin{tabular}{@{}llll}
\br
\#&Reaction & Process name &Threshold energy [eV]\\
\mr
1& $ \rm{e^{-}+He\longrightarrow e^{-} + He} $  &   Elastic scattering & 0 \\
2& $ \rm{e^{-}+He\longrightarrow e^{-} + He^\ast} $  & Electronic excitation: sum of triplets & 19.82 \\
3& $ \rm{e^{-}+He\longrightarrow e^{-} + He^\ast} $  & Electronic excitation: sum of singlets & 20.61 \\
4& $ \rm{e^{-}+He\longrightarrow 2e^{-} + He^+} $  &   Ionisation & 24.59 \\
\mr
5& $ \rm{e^{-}+N_{2} \longrightarrow e^{-} + N_{2}} $  &   Elastic scattering & 0 \\
6& $ \rm{e^{-}+N_{2}}  \longrightarrow \rm{e^{-} + N_{2}} (\textit{r}>0) $  &   Rotational excitation & 0.020 \\
7& $ \rm{e^{-}+N_{2}}  \longrightarrow \rm{e^{-} + N_{2}} (\textit{v}=1) $  &   Vibrational excitation & 0.290\\
8& $ \rm{e^{-}+N_{2}}  \longrightarrow \rm{e^{-} + N_{2}} (\textit{v}=1) $  &   Vibrational excitation & 0.291\\
9&$ \rm{e^{-}+N_{2}}  \longrightarrow \rm{e^{-} + N_{2}} (\textit{v}=2) $  &   Vibrational excitation & 0.590\\
10&$ \rm{e^{-}+N_{2}}  \longrightarrow \rm{e^{-} + N_{2}} (\textit{v}=3) $  &   Vibrational excitation & 0.880\\
11&$ \rm{e^{-}+N_{2}} \longrightarrow \rm{e^{-} + N_{2}} (\textit{v}=4) $  &   Vibrational excitation & 1.170\\
12&$ \rm{e^{-}+N_{2}}  \longrightarrow \rm{e^{-} + N_{2}} (\textit{v}=5) $  &   Vibrational excitation & 1.470\\
13&$ \rm{e^{-}+N_{2}}  \longrightarrow \rm{e^{-} + N_{2}} (\textit{v}=6) $  &   Vibrational excitation & 1.760\\
14&$ \rm{e^{-}+N_{2}} \longrightarrow \rm{e^{-} + N_{2}} (\textit{v}=7)$  &   Vibrational excitation & 2.060\\
15&$ \rm{e^{-}+N_{2}}  \longrightarrow \rm{e^{-} + N_{2}} (\textit{v}=8) $  &   Vibrational excitation & 2.350\\
16& $ \rm{e^{-}+N_{2}}  \longrightarrow \rm{e^{-} + N_{2}} (A~^3\Sigma_{\rm u}^+ ~$\textit{v}$ = 0-4) $  &   Electronic excitation & 6.170\\
17& $ \rm{e^{-}+N_{2}}  \longrightarrow \rm{e^{-} + N_{2}} (A~ ^3\Sigma_{\rm u}^+ ~$\textit{v}$ = 5-9) $  &   Electronic excitation & 7.000\\
18& $ \rm{e^{-}+N_{2}}  \longrightarrow \rm{e^{-} + N_{2}} (B~ ^3\Pi_{\rm g}) $  &  Electronic excitation & 7.350\\
19& $ \rm{e^{-}+N_{2}}  \longrightarrow \rm{e^{-} + N_{2}} (W~ ^3\Delta_{\rm u}) $  &  Electronic excitation & 7.360\\
20& $ \rm{e^{-}+N_{2}}  \longrightarrow \rm{e^{-} + N_{2}} (A~ ^3\Sigma_{\rm u}^+~ $\textit{v}$ > 10) $  &   Electronic excitation & 7.800\\
21& $ \rm{e^{-}+N_{2}}  \longrightarrow \rm{e^{-} + N_{2}} (B$'$~ ^3\Sigma_{\rm u}^-) $  &   Electronic excitation & 8.160\\
22& $ \rm{e^{-}+N_{2}}  \longrightarrow \rm{e^{-} + N_{2}} (a$'$~ ^1\Sigma_{\rm u}^-) $  &   Electronic excitation & 8.400\\
23& $ \rm{e^{-}+N_{2}}  \longrightarrow \rm{e^{-} + N_{2}} (a~ ^1\Pi_{\rm g}) $  &   Electronic excitation & 8.550\\
24& $ \rm{e^{-}+N_{2}}  \longrightarrow \rm{e^{-} + N_{2}} (w~ ^1\Delta_{\rm u}) $  &   Electronic excitation & 8.890\\
25& $ \rm{e^{-}+N_{2}}  \longrightarrow \rm{e^{-} + N_{2}} (C~ ^3\Pi_{\rm u}) $  &   Electronic excitation & 11.03\\
26& $ \rm{e^{-}+N_{2}}  \longrightarrow \rm{e^{-} + N_{2}} (E~ ^3\Sigma_{\rm g}^+) $  &   Electronic excitation & 11.88\\
27& $ \rm{e^{-}+N_{2}}  \longrightarrow \rm{e^{-} + N_{2}} (a$''$~ ^1\Sigma_{\rm u}^+) $  &   Electronic excitation & 12.25\\
28& $ \rm{e^{-}+N_{2}} \longrightarrow \rm{e^{-}} + N + N $  &  Dissociation & 13.00\\
29& $ \rm{e^{-}+N_{2}} \longrightarrow 2\rm{e^{-}} + N_2^+ $  &  Ionisation & 15.60\\
\mr
30& $\rm {He (2^1{\rm S},2^3{\rm S}) + {\rm N}_2 \longrightarrow {\rm He} (1^1{\rm S}) + {\rm N}_2^+ + {\rm e}^-} $ & Penning ionisation & - \\
31& $\rm {He^+ + {\rm He}  \longrightarrow {\rm He}^+ + {\rm He}}$ & Elastic scattering & - \\
32& $\rm {He^+ + {\rm He} + {\rm He} \longrightarrow {\rm He}_2^+ + {\rm He}}$ & Ion conversion & - \\
33& $\rm {He_2^+ + {\rm He} \longrightarrow {\rm He}_2^+ + {\rm He}}$ & Elastic scattering & - \\
34& $\rm {N_2^+ + {\rm He} \longrightarrow {\rm N}_2^+ + {\rm He}}$ & Elastic scattering & - \\
\mr
35& $\rm {He(2^1P) \longrightarrow He(1^1S) + photon}$ & Resonance photon emission & NA\\
35& $\rm {He(1^1S) + photon \longrightarrow He(2^1P)}$ & Resonance photon absorption & NA \\
\br
\end{tabular}
\end{table}

\vspace{0.5cm}

{\it Electrons} may collide with the background gas, consisting of He atoms and N$_2$ molecules. 
For electron -- He atom collisions we use the cross sections from \cite{he-cs}, while for electron -- N$_2$ collisions we use the cross section set of \cite{n2-cs} (that was largely based on the Siglo set, now accessible via LxCat \cite{siglo}). When electrons collide with He atoms, a significant part of the excitation is assumed to lead to the formation of (singlet, 2$^1$S and triplet, 2$^3$S) metastable atoms and atoms in the lowest resonant state (2$^1$P). While the exact rates of creation and the densities of these specific states could only be obtained by developing a full collisional -- radiative model, this is beyond the scope of this work. Instead of this, we adopt the following approximations: (i) 50\% of the total excitations leading to He singlet states are assumed to result in the formation of atoms in the 2$^1$S singlet metastable state and the other 50\% leads to the formation of atoms in the 2$^1$P resonant state. This is justified by the relatively low rate of the direct decay of higher excited states to the ground state and the importance of cascade transitions that lead to the population of the 2$^1$S and 2$^1$P states. (ii) Regarding excitation processes in the triplet system, 50\% is assumed to lead to the formation of 2$^3$S metastable atoms. According to these simplifications we need to trace only the resonance photons emitted by the He atoms in the 2$^1$P state. (We note that the results of \cite{Fierro} indicated a strong dominance of the 2$^1$P $\rightarrow$ 1$^1$S radiation, over the $n^1$P $\rightarrow$ 1$^1$S transitions, with $n > 2$, see figure 9 of \cite{Fierro}.) The method of following the VUV resonance photons will be outlined in section 2.2. 

The electron--N$_2$ molecule collisions include elastic scattering, rotational, vibrational and electronic excitation and ionization. Among the "products" of these reactions only N$_2^+$ ions are followed in the simulation, as the concentration of N$_2$ does not exceed 1\% in the background gas. 

All types of electron-atom/molecule collisions are assumed to result in isotropic scattering.

\vspace{0.5cm}

{\it Helium metastable atoms} contribute efficiently to the ionization, when N$_2$ is present in the background gas, via the Penning ionization process \cite{Niemi}: 
\begin{equation}
{\rm He} (2^1{\rm S},2^3{\rm S}) + {\rm N}_2 \rightarrow {\rm He} (1^1{\rm S}) + {\rm N}_2^+ + {\rm e}^-.
\label{eq:penning}
\end{equation} 
The rate of these reactions is $k_{\rm P} = 5.0 \times 10^{-17}$ m$^3$ s$^{-1}$, as given in \cite{Brok,Sakiyama}. In our particle-based approach a lifetime $\tau_{\rm m}$ is assigned to each metastable atom upon its "birth" at time $t =t_{\rm m}$, according to 
\begin{equation}
\tau_{\rm m} = - \frac{1}{k_{\rm P} n_{\rm N_2}} \ln (1-R_{01}).
\end{equation}
Using random numbers ($R_{01}$) with uniform distribution over the $[0,1)$ interval this formula generates samples of $\tau_{\rm m}$ with an exponential probability density distribution. In the following time steps of the simulation the validity of $t > t_{\rm m} + \tau_{\rm m}$, where $t$ is the actual time, is evaluated. The metastable atom is "destroyed" and an ${\rm N}_2^+$ ion is created if the inequality is found to hold. The thermal motion (diffusion) of metastable atoms is neglected because of the short time scales considered.

\vspace{0.5cm}

{\it Helium atomic ions} experience elastic collisions with He atoms of the background gas; we follow the approach of Phelps and partition elastic collisions into an isotropic channel and a backward scattering part \cite{Phelps}. Due to the low concentration of nitrogen, we disregard He$^+$ + N$_2$ collisions. 

\vspace{0.5cm}

{\it Helium molecular ions} are created efficiently via the three-body ion conversion process
\begin{equation}
{\rm He}^+ + {\rm He} + {\rm He} \rightarrow {\rm He}_2^+ + {\rm He},
\label{eq:conv}
\end{equation} 
at the elevated pressure (1 bar) considered here. The rate of this process, $k_{\rm conv} = 1.1 \times 10^{-43} {\rm m}^6 {\rm s}^{-1}$, is taken from \cite{Brok,Sakiyama}. Similarly to the Penning ionization process we assign a lifetime $\tau_{\rm c}$  to the He$^+$ ions (upon their "birth") as 
\begin{equation}
\tau_{\rm c} = - \frac{1}{k_{\rm conv} n_{\rm He}^2} \ln (1-R_{01}).
\end{equation}
Conversion of He$^+$ to He$_2^+$ is executed after this lifetime. The motion of the ${\rm He}_2^+$ ions is limited by their collisions with the background He atoms, for which the Langevin cross section is used:
\begin{equation}
\sigma_{\rm L} = \sqrt{\frac{\pi \alpha q^2}{\varepsilon_0 \mu}} \frac{1}{g},
\end{equation}
where $q$ is the elementary charge, $\alpha$ is the polarizibility of He atoms, $\mu$ is the reduced mass of the projectile (He$_2^+$) and target (He) species, and $g$ is the relative velocity of the collision partners. (Note that as the collision probability $P$ involves the product of $g\,\sigma_{\rm L}$ (see eq. (\ref{eq:pcoll1})), $P$ is independent of $g$. N$_2$ molecules are not considered as targets for ${\rm He}_2^+$ ions due to their low concentration.

\vspace{0.5cm}

{\it Nitrogen molecular ions} are also assumed to interact with He atoms only. This interaction is also described by a Langevin cross section, similar as above.

\subsection{Photon transport}

In our model photons are treated as discrete particles that move with the speed of light. They originate from emission events from the lowest resonant state of He, which is the only state considered as a source of VUV radiation. This state (He 2$^1$P) has a lifetime of $\tau_0 = 0.56$ ns \cite{Zitnik} and the transition to the ground state has a central wavelength of $\lambda_0 = 58.4334$ nm. The wavelength of the radiation actually emitted by an atom, is, however, usually different from this value due to line broadening mechanisms, which are of homogeneous and inhomogeneous characters (see e.g. \cite{Kunze}). The finite lifetime of the excited state results in {\it natural broadening} and the interaction of the radiating atom with the surrounding atoms results in {\it pressure broadening}. These mechanisms broaden the spectrum in a homogeneous way, creating a Lorentzian line shape. (Another contribution to homogeneous broadening may be due to the Stark-effect, this, however, is expected to be small compared to the above mentioned mechanisms, for the conditions considered here \cite{Dimitrijevic}.) The {\it Doppler mechanism} is responsible for causing inhomogeneous broadening that gives rise to a Gaussian line shape. The ensemble of radiated photons will follow, as a consequence of these three mechanisms, a Voigt-type spectral line shape that is the convolution of Lorentzian and Gaussian line shapes. The propagating photons can be absorbed by ground-state He atoms according to photoabsorption cross sections that depends on their actual wavelength (see below). The absorption events "create" a He atom in the 2$^1$P state that will, subsequently, radiate after a random delay corresponding to the natural lifetime. The above processes are repeated in the simulation until the photon reaches of the electrodes or until the simulation terminates at a defined time.

The broadening mechanisms are considered in our simulations by Monte Carlo approach, through generating random samples. For the homogeneous part of line broadening, random samples are taken of the corresponding Lorentzian distribution, i.e., random wavelengths to each radiated photon are assigned according to \cite{Fierro}:
\begin{equation}
\lambda_1 = \tan \big[(R_{01}-0.5) \pi \big] \Delta \lambda_{\rm L} + \lambda_0,
\label{eq:L1}
\end{equation}
where $\Delta \lambda_{\rm L} = \Delta \lambda_{\rm n}+\Delta \lambda_{\rm p}$ is the full width at half maximum (FWHM) of the Lorentzian curve originating from the contributions of natural ($\Delta \lambda_{\rm n}$) and pressure ($\Delta \lambda_{\rm p}$) broadening components. The latter are given by \cite{Kunze}:
\begin{equation}
\Delta \lambda_{\rm n} = \lambda_0^2 \frac{A_{21}}{2 \pi c}
\end{equation}
and 
\begin{equation}
\Delta \lambda_{\rm p} = 9 \times 10^{-16} \sqrt{\frac{g_1}{g_2}} f_{12}\lambda_0^3 n_{\rm g}.
\end{equation}
Here $c$ is the speed of the light, $A_{21}= 1 / \tau_0$ is the Einstein coefficient for spontaneous emission, $n_{\rm g}$ is the background gas density, $g_1$ and $g_2$ are, respectively, the statistical weights of the ground state and the 2$^1$P state, and $f_{12}$ is the {\it oscillator strength}:
\begin{equation}
f_{12} = \frac{1}{3} \frac{A_{21}}{\gamma_{\rm cl}} \frac{g_2}{g_1}, 
\end{equation}
which involves the classical decay rate
\begin{equation}
\gamma_{\rm cl}=\frac{q^2 \omega_0^2}{6 \pi \varepsilon_0 m_{\rm e} c^3},
\end{equation}
where $m_{\rm e}$ is the mass of the electron.

\begin{figure}[ht ]
\begin{center}
\includegraphics[width =0.52\textwidth]{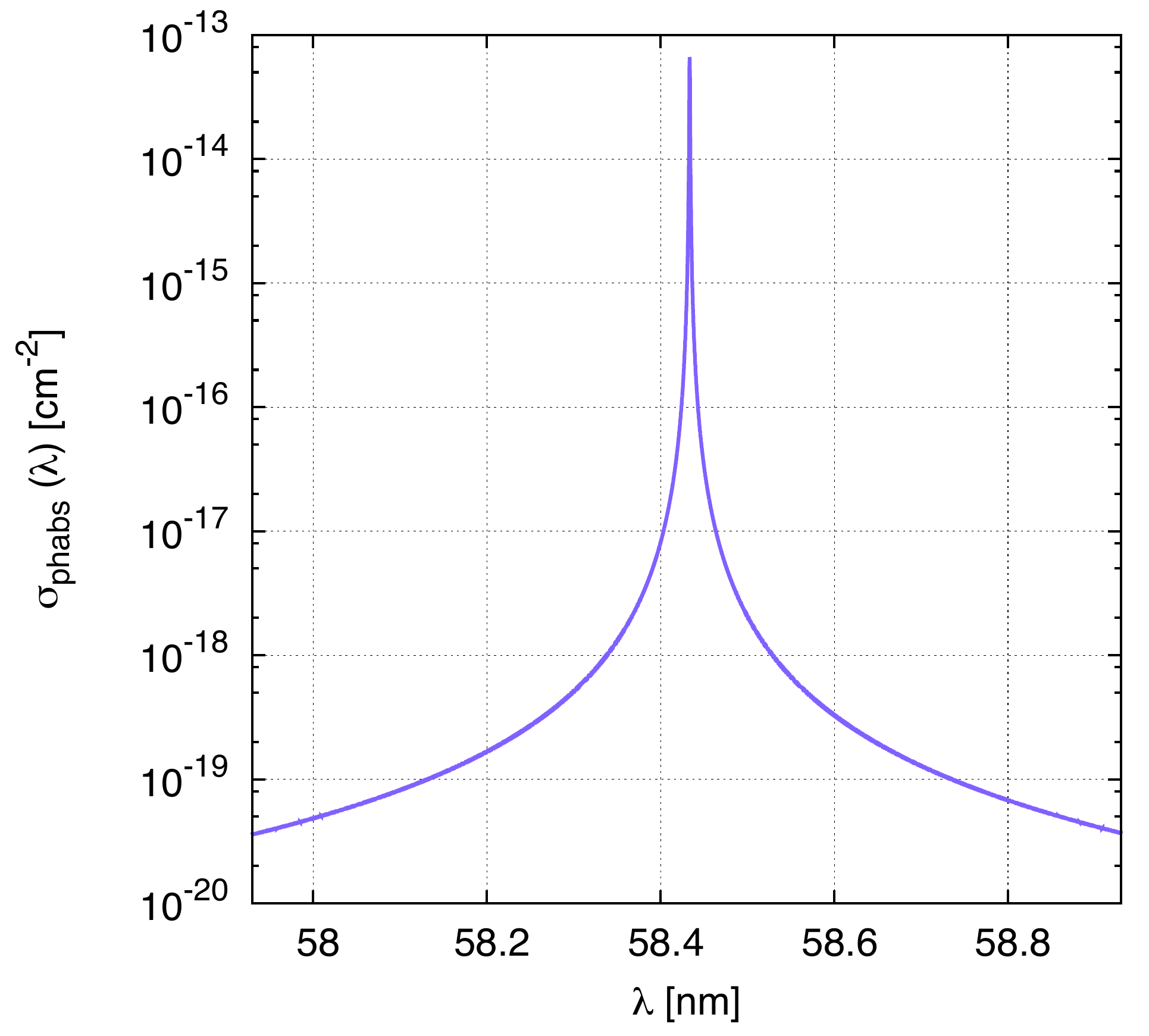}
\caption{Photoabsorption cross section for the He 2$^1$P $\rightarrow$ 2$^1$S resonance radiation.}
\label{fig:phabs}
\end{center}
\end{figure}

The wavelength assigned according to eq. (\ref{eq:L1}) to each radiated photon is modified by the Doppler shift that is caused by the thermal motion of the radiating atom. In this procedure, first, a random velocity vector, ${\bf v}_{\rm A}$, for the (radiating) atom is sampled from the Maxwell-Boltzmann distribution of the background atoms. Secondly, a random direction, via a unit vector $\hat{\bf e}_{\rm ph}$, to the radiated photon is assigned. With these quantities given, the photon is radiated with a wavelength \cite{Fierro}:
\begin{equation}
\lambda = \frac{(c + \hat{\bf e}_{\rm ph}\cdot {\bf v}_{\rm A}) ~\lambda_1}{c}.
\end{equation}
The probability of the absorption of the photons is computed using eq.(\ref{eq:pcoll1}), with $n_{\rm targ} \leftarrow n_{\rm He}$, $\sigma_{\rm proj, tot} \leftarrow \sigma_{\rm phabs}$, and $\Delta t \leftarrow \Delta t_{\rm ph}$ replacements, where $\Delta t_{\rm ph}$ is the time step for photons (see below) and $\sigma_{\rm phabs}$ is the photoabsorption cross section. The latter is given as 
\begin{equation}
\sigma_{\rm phabs} = \frac{g_2}{g_1} \frac{\lambda_0^4 A_{21}}{8 \pi c} V(\lambda),
\label{eq:phabs}
\end{equation}
where $V$ is the Voigt-profile function computed by the algorithm given in \cite{Voigt} based on the expression
\begin{eqnarray}
V(\lambda) = \frac{Y}{\pi} \int_{-\infty}^{\infty} \frac{\exp(-t^2)}{Y^2 + (X-t)^2}
{\rm d}t,\nonumber \\
X = \frac{\lambda-\lambda_0}{\Delta \lambda_{\rm G}} 2 \sqrt{\ln2},~~~~
Y = \frac{\Delta \lambda_{\rm L}}{\Delta \lambda_{\rm G}} \sqrt{\ln2} \nonumber
\end{eqnarray}
and it satisfies $\int V(\lambda) {\rm d} \lambda = 1$. Here $\Delta \lambda_{\rm G}$ is the line width due to Doppler broadening. The cross section computed according to (\ref{eq:phabs}) is shown in figure~\ref{fig:phabs}.

The cross section (\ref{eq:phabs}) and probability of absorption during a given time step is computed by taking a randomly chosen (potentially absorbing) atom from the background gas with velocity ${\bf v}_{\rm B}$, and using the wavelength 
\begin{equation}
\lambda^\ast = \frac{(c - \hat{\bf e}_{\rm ph} \cdot {\bf v}_{\rm B}) \lambda}{c},
\end{equation}
that takes into account the Doppler shift caused by the motion of the potentially absorbing atom. Whenever an absorption event takes place, a (random) lifetime to the excited state is assigned as
\begin{equation}
\tau_{\rm exc} = - \tau_0 \ln(1-R_{01}).
\end{equation}
Emission events, of which the details were given above, are initiated when the lifetime is passed. 

\subsection{Stability constraints and the timing of the simulation}

The stability and accuracy of PIC/MCC simulations is guaranteed only when the relevant criteria are fulfilled. These include the requirements that the computational grid resolves the Debye length, the discrete time step for a given species resolves its plasma frequency and another condition, known as the Courant condition, ensures that charged particles do not move more than a grid division during a time step. The spatial grid uses $N_{\rm grid}$ = 800 points, providing a resolution that is in the order of the smallest Debye length.

Additionally, the time step of a given species should be short enough to ensure that the probability of  multiple collisions during a time step is negligible (otherwise some of the collisions will be missed in the simulation). Due to the very high collisionality of the plasma at atmospheric pressure this last condition defines the upper bound for the time step. 

To optimize the code different time steps for the various species are used. In practice, a collision probability 
\begin{equation}
P = 1- \exp(-\nu \Delta t)
\end{equation}
that amounts a few percent, over the whole relevant domain of collision frequencies ($\nu$) is usually allowed. Here we set the limit at 10\%, i.e. set the time step for a given species according to 
\begin{equation}
P = 1- \exp(-\nu_{\rm max} \Delta t) = 0.1~~\rightarrow~~\Delta t \cong 0.105 / \nu_{\rm max}.
\end{equation}
Here, $\nu_{\rm max}$ is the maximum collision frequency that can be derived from the ensemble of cross sections of the given species. 

Among the charged species, electrons have the highest collision frequency, which limits their time step to $\Delta t_{\rm e} = 4.5 \times 10^{-14}$ s. The time steps for the different ionic species are set, considering their respective maximum collision frequencies, according to $\Delta t_{\rm He^+} = 10 \, \Delta t_{\rm e}$ and $\Delta t_{\rm He_2^+} = \Delta t_{\rm N_2^+} = 100 \,\Delta t_{\rm e}$. The time step for electrons is chosen to represent the basic "tick" of the simulation, the ions are traced only in fewer number of ticks, but with correspondingly longer time steps. This procedure is known in the literature as "sub-cycling".

For the resonance photons the high photoabsorption cross section poses a significantly lower time step, as compared to electrons. Thus, oppositely to ions: we use "super-cycling" for the photons, in each tick of the simulation the photons are traced with $\Delta t_{\rm ph} = \Delta t_{\rm e} / 10000$, meaning that the time step for photons is $\Delta t_{\rm ph} = 4.5 \times 10^{-18}$ s. While this time step seems prohibitively small for a simulation that needs to run to about 100 ns, as the number of photons turns out to be significantly smaller than the number of charged particles, the simulation runs are actually feasible even at this  extremely small required time step.

Electron emission from the electrodes is assumed to be dominated by ejection due to the impact of ions and VUV photons.  We do not consider secondary electron emission due to electron impact, but include the reflection of impinging electrons (in a simplified way) with a constant probability of $P_{\rm e} = 0.2$ \cite{Kollath}. 

The simulations start with seeding electrons and He$_2^+$ ions within the discharge gap with an equal number ($N_0$), with a uniform spatial distribution. These charged particles emulate the charges remaining from a previous pulse. (We assume a repetitive operation of the system, which we are, however, not able to follow due to the very intensive computational needs that prohibit simulation of low duty cycle settings, with typically $\sim$ kHz repetition rates.) The corresponding initial charged particle density is defined via setting the weight, $W$, of the superparticles.

The effect of recombination processes between electrons and He$^+$ and N$_2^+$ ions was tested and found to have negligible influence on the charged particle densities within the time domain considered. Therefore, these processes, as they generate an unnecessary computational overhead, are not included in the simulations presented here. Recombination processes are, however, important on the time scale of the typical interval between the voltage pulses ($\sim$ ms) in experiments, where the density of "remaining" charges is defined by the decay rates due to recombination and ambipolar diffusion. The actual spatial distribution of the charged species, in the typical range of experimental conditions, may have some influence on the discharge initiation, e.g., (i) the temporal distribution of the flux of the "initial" ions at the cathode influences the development of electron avalanches and (ii) a significant density may distort the field distribution from the beginning of the excitation pulse. The first effect is expected to be less influential on the results than uncertainties of the surface coefficients, and, regarding the second effect, we can assume that the "initial" charge density is low enough not to create a significant distortion of the field after the application of the excitation pulse that breaks quasi-neutrality established in the afterglow periods.

\section{Results}

Below we present the simulation results for $p$ = 1 bar pressure and $L$ = 1 mm electrode gap. The density of the background gas is computed by taking the gas temperature to be 300 K. Although heating of the gas occurs during the pulses (see e.g. \cite{Uwe}), we do not expect this to result in a significant depletion of the density in an experimental system over the short time scales considered here.
The computations are carried out at the following values of the secondary electron emission coefficients:
$\gamma_{\rm He^+} = 0.15$, $\gamma_{\rm He_2^+} = 0.1$, $\gamma_{\rm N_2^+} = 0.05$, and $\gamma_{\rm ph} = 0.1$, where the last value refers to the VUV resonance photons radiated from the He 2$^1$P state. These ion-induced yields of the helium ionic species are somewhat lower than those adopted in \cite{Sakiyama}, for He$^+$ we base our choice on the findings of \cite{donko2006,Kutasi}, and for He$_2^+$ we apply the findings of \cite{PH} that their yield is $\approx$ 60\% of the yield of atomic ions. These coefficients are known, of course, to have a large uncertainty related to the properties of the electrode surfaces (see, e.g., \cite{PP99}). These uncertainties are, however, not expected to change our results and conclusions dramatically as long as reasonable values for the $\gamma$-coefficients are used. 

In the simulations we assume an $A$ = 1 cm$^2$ electrode area. The number of superparticles representing electrons and He$_2^+$ ions seeded at the beginning of the simulations is $N_0 = 1.5 \times 10^4$ (for both species). As the discharge volume considered is $A L$ = 0.1 cm$^{3}$, the density of the above species is $n_0 = N_0 W /  A L$. Here, $W$ is the weight of the superparticles (equal to the number of actual particles represented by one superparticle).

\begin{figure}[ht ]
\begin{center}
\small (a) ~~~~~~~\includegraphics[width =0.52\textwidth]{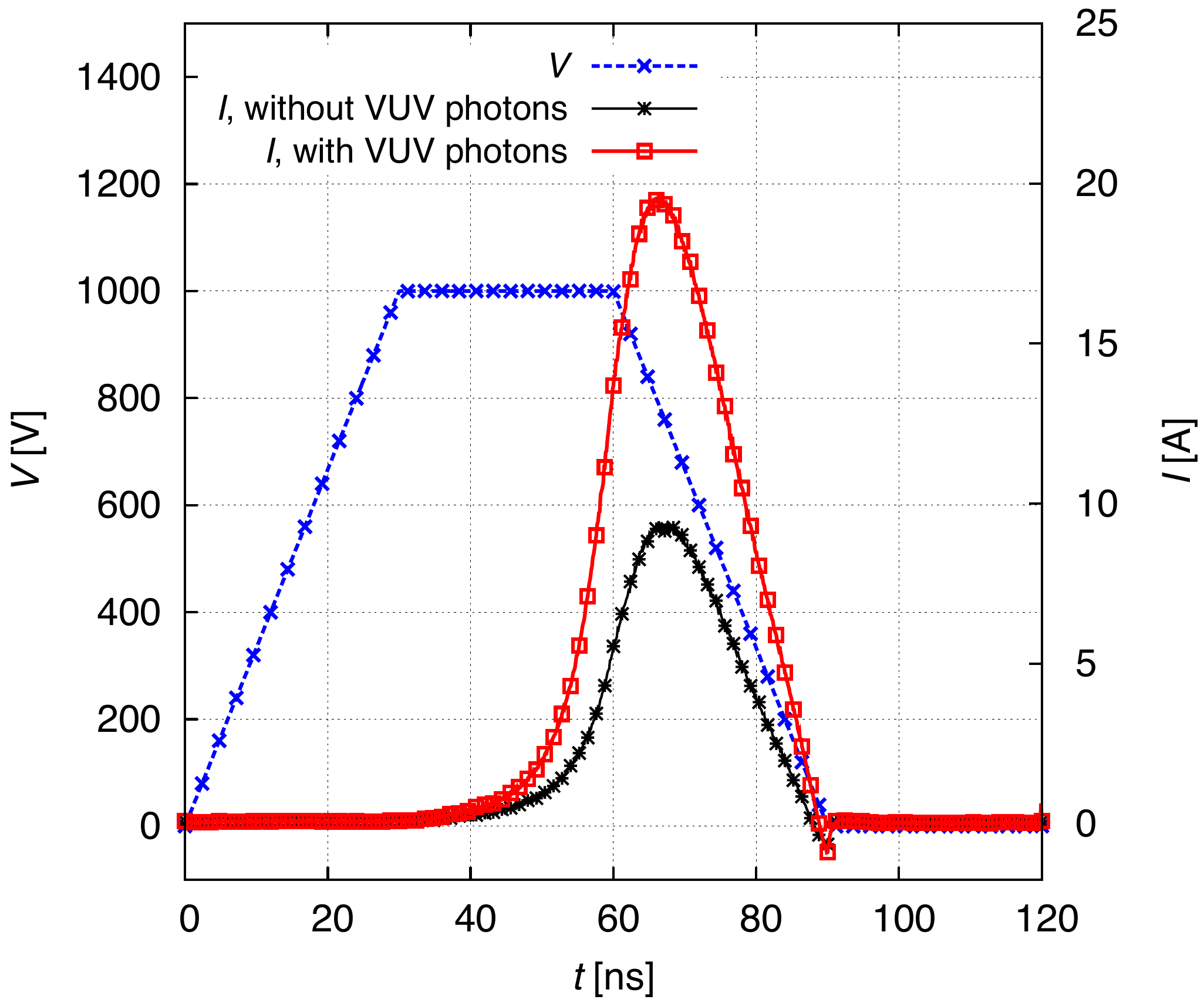}\\
\small (b) \includegraphics[width =0.47\textwidth]{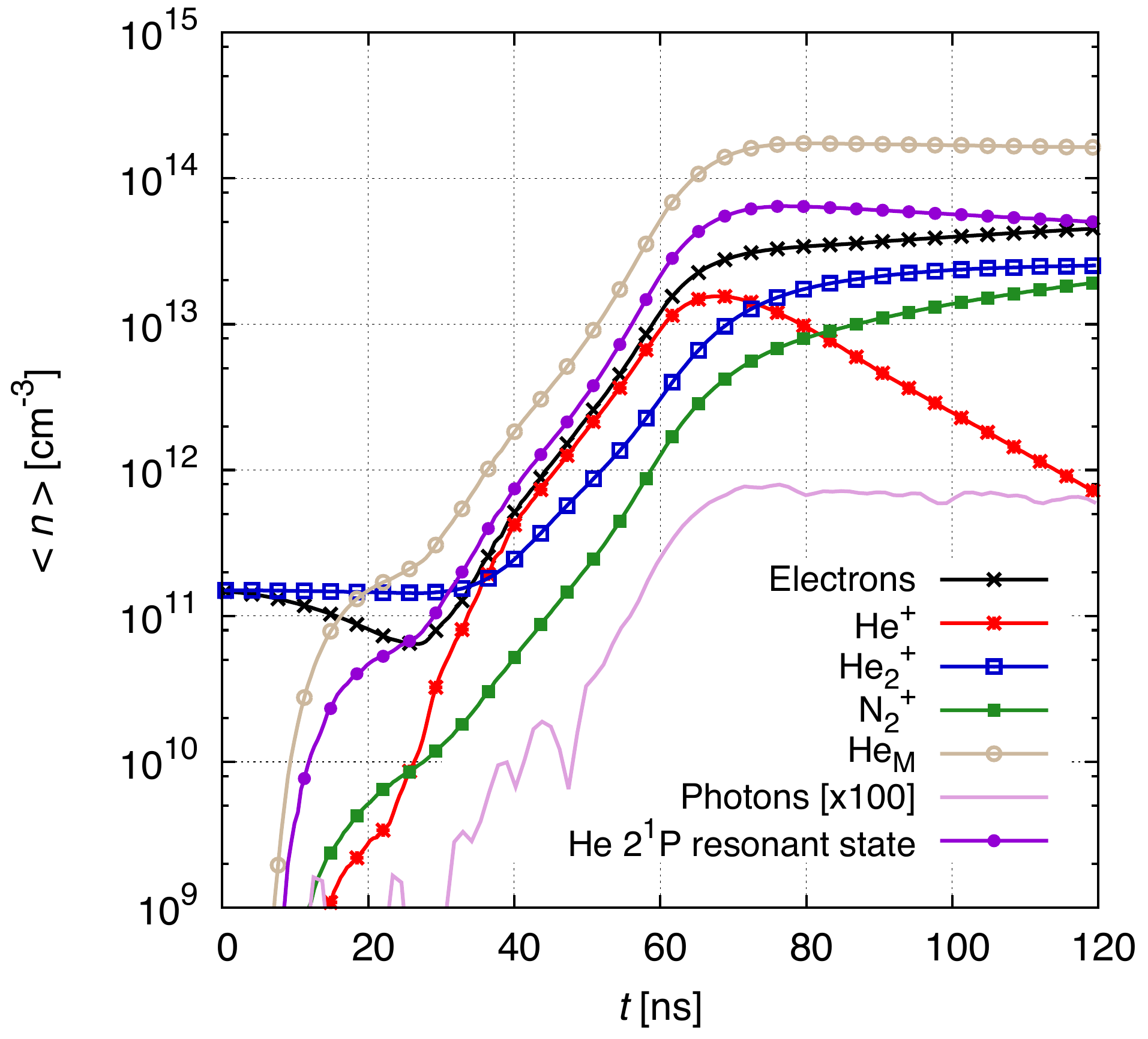}
\caption{(a) Excitation voltage waveform (blue dashed line) and discharge current obtained in the simulations including (red solid line) and excluding (black solid line) VUV photons, for He + 0.1\% N$_2$,  $p$ = 1 bar, $L$ = 1 mm, and $n_0 = 1.5 \times 10^{11}$ cm$^{-3}$. The peak of the voltage pulse and the duration of the plateau are $V_0$ = 1000 V and $\tau_{\rm p}$ = 30 ns, respectively. These parameter settings correspond to the "base case". (b) Time-dependence of the spatially averaged densities of the species / states considered in the simulation for the same set of conditions. (Note that the curve for the VUV photons is 100$\times$ multiplied.)}
\label{fig:currents1}
\end{center}
\end{figure}

Simulations have been carried out with excitation voltage pulses with different lengths and peak values ($V_0$), but the same model trapeziodal excitation waveform is always used: the rise time ($\tau_\uparrow$), the length of the plateau ($\tau_{\rm p}$) and the fall time ($\tau_\downarrow$) of the pulses are the same ($\tau_\uparrow=\tau_{\rm p} = \tau_\downarrow$). For an example, see figure~\ref{fig:currents1}, where the dashed curve shows such a voltage pulse with $V_0$ = 1000 V peak value and $\tau_{\rm p}$ = 30 ns plateau length. These values of peak voltage and plateau length will be used as the "base case" in the following, for which detailed illustration and analysis of the discharge physics will be given in section \ref{sec:res1}. For the base case, moreover, the initial density of charged particles is set to $n_0 = 1.5 \times 10^{11}$ cm$^{-3}$ (taking $W = 10^6$) and the composition of the gas is set to He + 0.1\% N$_2$. The effects of the variation of these parameters and the amount of the N$_2$ admixture is investigated in section 3.2.

\subsection{Discharge properties in the base case}

\label{sec:res1}

At the parameter settings corresponding to the base case, a current pulse having a peak value of 20 A and a FWHM of $\sim$ 20 ns is generated by the high voltage pulse excitation, as it can be seen in figure \ref{fig:currents1}(a), when the photon-induced emission of secondary electrons from the electrodes is considered. For the specific conditions the current pulse peaks during the falling edge of the excitation signal and terminates, following a small negative peak, at the same time as the excitation pulse. The reason for the creation of the negative current is the appearance of a large displacement current, as it will be shown later. When the effect of VUV photons is disregarded, a factor of two lower current is obtained from the simulations (for this base case). This points out the importance of the photoemission process in the quantitative characterization of the discharge plasma. The contribution of this effect to secondary electron emission will be compared later on to the contributions of the other species.    

\begin{figure}[h!]
\begin{center}
\small (a) \includegraphics[width =0.5\textwidth]{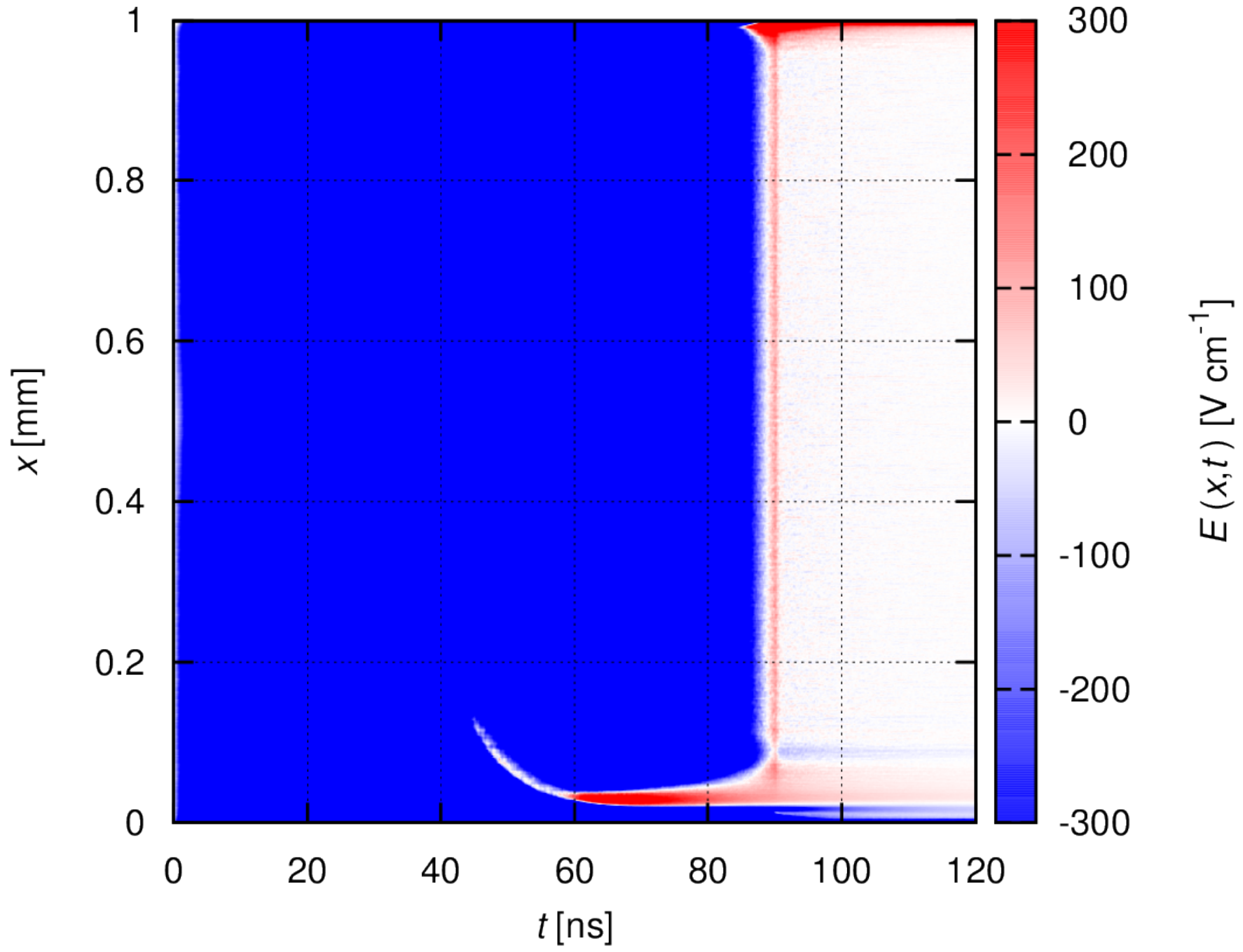}\\
\small (b) \includegraphics[width =0.5\textwidth]{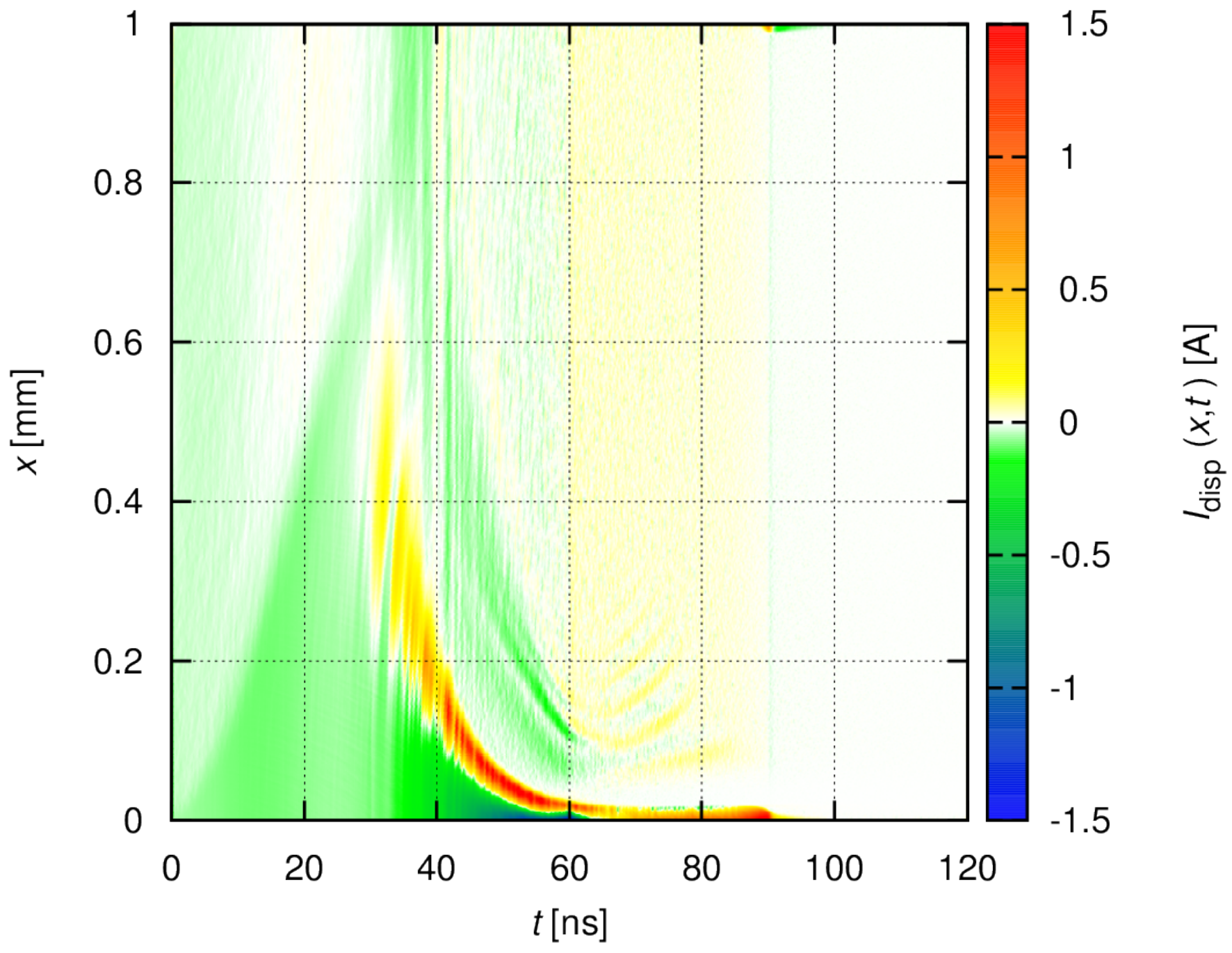}
\caption{Spatio-temporal distributions of the (a) electric field and (b) displacement current, for the base conditions (He + 0.1\% N$_2$, $p$ = 1 bar,  $L$ = 1 mm, $V_0$ = 1000 V, $\tau_{\rm p}$ = 30 ns,  $n_0 = 1.5 \times 10^{11}$ cm$^{-3}$, VUV photons considered). The distribution of the electric field is shown only at low magnitudes, the color scale is saturated outside the $|E| > 300$ V\,cm$^{-1}$ domain. Note the regions with "reversed field", i.e. $E>0$. The cathode is situated at $x=0$ mm, the anode is at $x=1$ mm.}
\label{fig:field-disp}
\end{center}
\end{figure}

The shape of the current pulse in figure \ref{fig:currents1}(a) can be understood by observing the evolution of the densities of the different charged species and excited states considered in the simulation. At early times, before $\sim$ 30 ns, we observe in figure \ref{fig:currents1}(b) that the electron density decays from the initial value, due to the effect of the continuously increasing electric field that sweeps electrons towards the anode. In this domain the electric field is nearly homogeneous, due to the relatively low charged particle density. A significant increase of the density of excited atoms (both in the metastable states and in the lowest resonant state) already takes place in this time domain, while a rapid growth of the electron and (He$^+$ and N$_2^+$) ion densities is found only beyond $\sim$ 30 ns, when the excitation voltage pulse reaches its plateau. The mean electron density increases during the next 30 ns by almost three orders of magnitude. During the falling edge of the voltage pulse (60 ns $\leq t \leq$ 90 ns) the excitation and ionization of helium becomes less significant. Consequently, the density of the 2$^1$P state starts to decrease slightly. A much faster decay of the He$^+$ density is observed at the same time, due to the conversion reaction (\ref{eq:conv}) that results in an increasing He$_2^+$ density. The high density of metastable atoms serves as a supply for the creation of N$_2^+$ ions (via the process (\ref{eq:penning})) even after the termination of the voltage pulse. This way the N$_2^+$ ion density reaches its maximum well beyond the termination of the voltage pulse. We note that the density of photons is approximately four orders of magnitude lower as compared to the density of the 2$^1$P-state atoms. This is due to the very short "flights" of the VUV photons between the emission and reabsorption events, compared to the lifetime of the 2$^1$P excited state. The total number of charged (super)particles followed in the simulation reaches about 10 million at the end of the run. This magnitude of the number of particles was found to generate data with sufficient statistics and results in good reproducibility.  

The development of the discharge plasma can also be followed by monitoring the spatio-temporal distributions of the electric field and the displacement current, as shown in figures \ref{fig:field-disp}(a) and (b), respectively. The range of the electric field is restricted here to small magnitudes, in order to be able to see finer, lower-field structures. During the increase of the voltage the electric field becomes distorted due to the accumulation of space charges. At times $t \geq 45$ ns a narrow low-field region develops that is a feature of sheath formation after breakdown. This region moves closer towards the cathode during the period of constant voltage due to increasing plasma density. This is followed by the development of a field reversal during a period of decreasing voltage starting at 60 ns. This field reversal develops due to an excess of positive space charge. The positive space charge in the region  would be sufficient to shield the previously applied high voltage through the sheath in front of the electrode. During the decrease of this voltage the collision limited mobility of electrons does not allow an instant adjustment to move the sheath edge - resulting in a localised excess of positive space charge and an associated transient field reversal. At the termination of the voltage pulse a narrow reversed-field region appears along the whole gap, and, subsequently most of the gap becomes field-free, but near the cathode a triple layer forms. The main feature seen in the distribution of the  displacement current (figure \ref{fig:field-disp}(b)) is also related to the formation of the cathode region. A strong displacement current pulse develops at $t \cong$ 90 ns, at the termination of the excitation voltage pulse. 

\begin{figure}[h!]
\begin{center}
\includegraphics[width =0.5\textwidth]{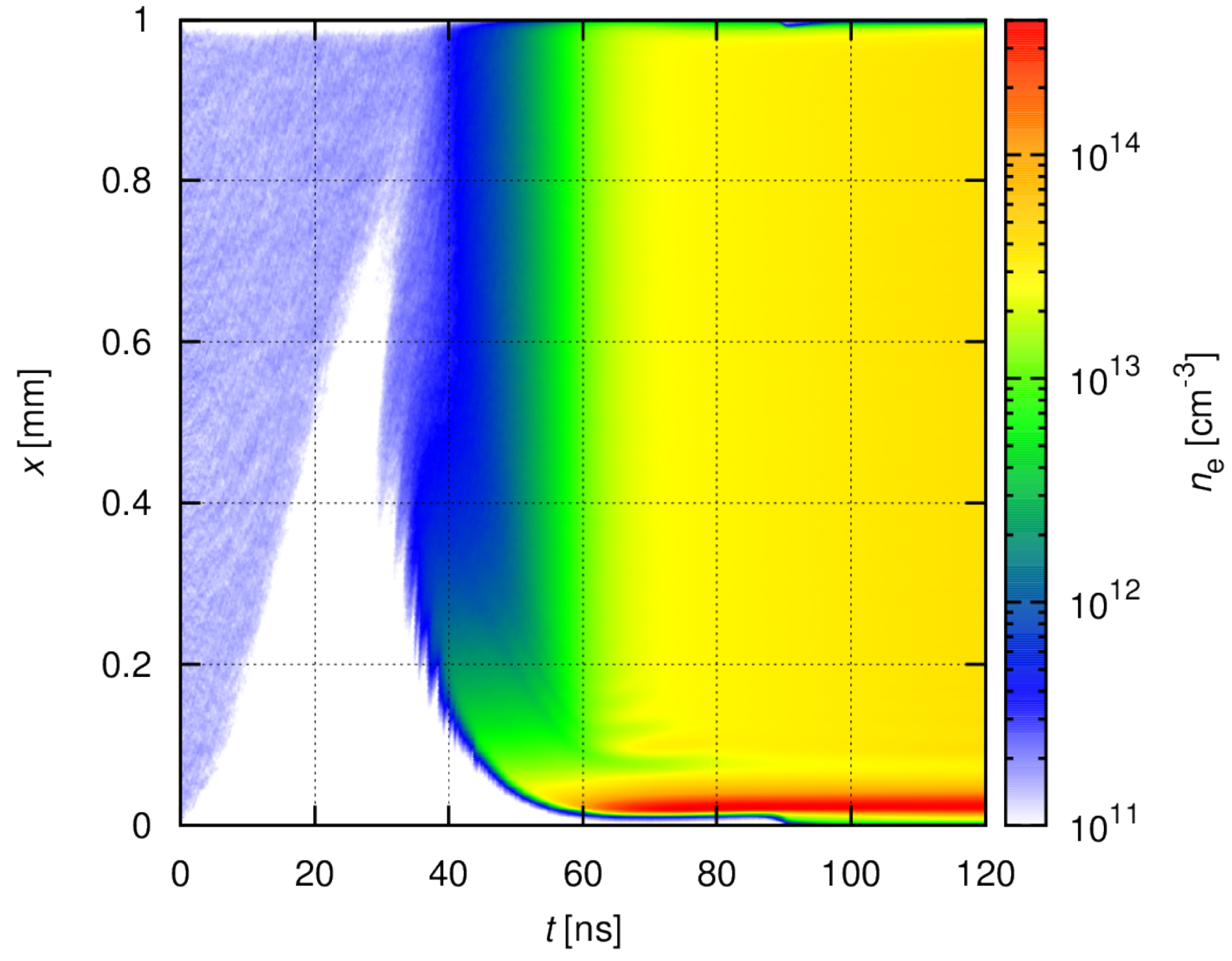}
\caption{(a) Spatio-temporal distribution of the electron density for the base conditions: He + 0.1\% N$_2$, at $p$ = 1 bar,  $L$ = 1 mm, $V_0$ = 1000 V, $\tau_{\rm p}$ = 30 ns, and $n_0 = 1.5 \times 10^{11}$ cm$^{-3}$, with VUV photons considered.}
\label{fig:edensity}
\end{center}
\end{figure}

\begin{figure}[ht ]
\begin{center}
\small (a) \includegraphics[width =0.5\textwidth]{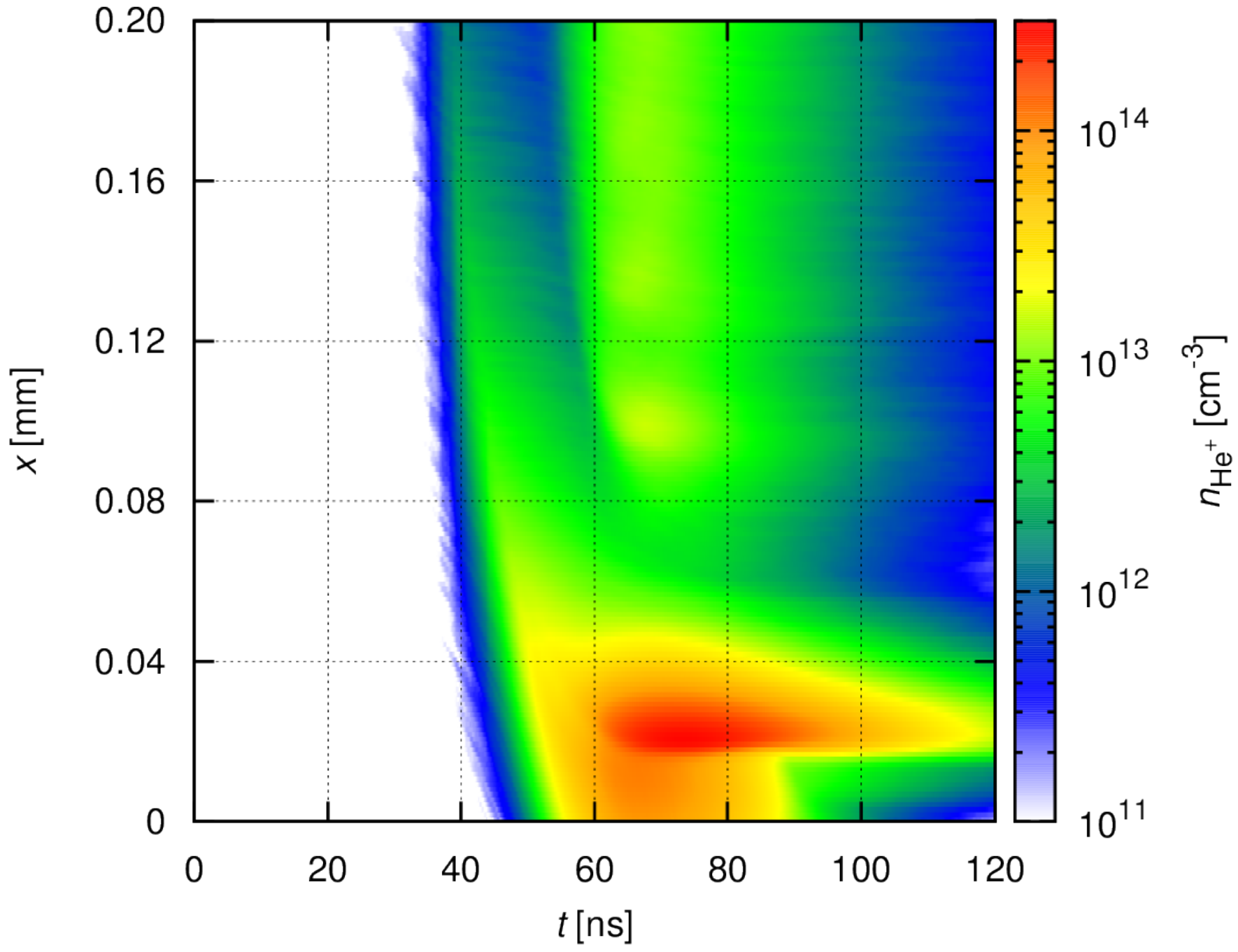}\\
\small (b) \includegraphics[width =0.5\textwidth]{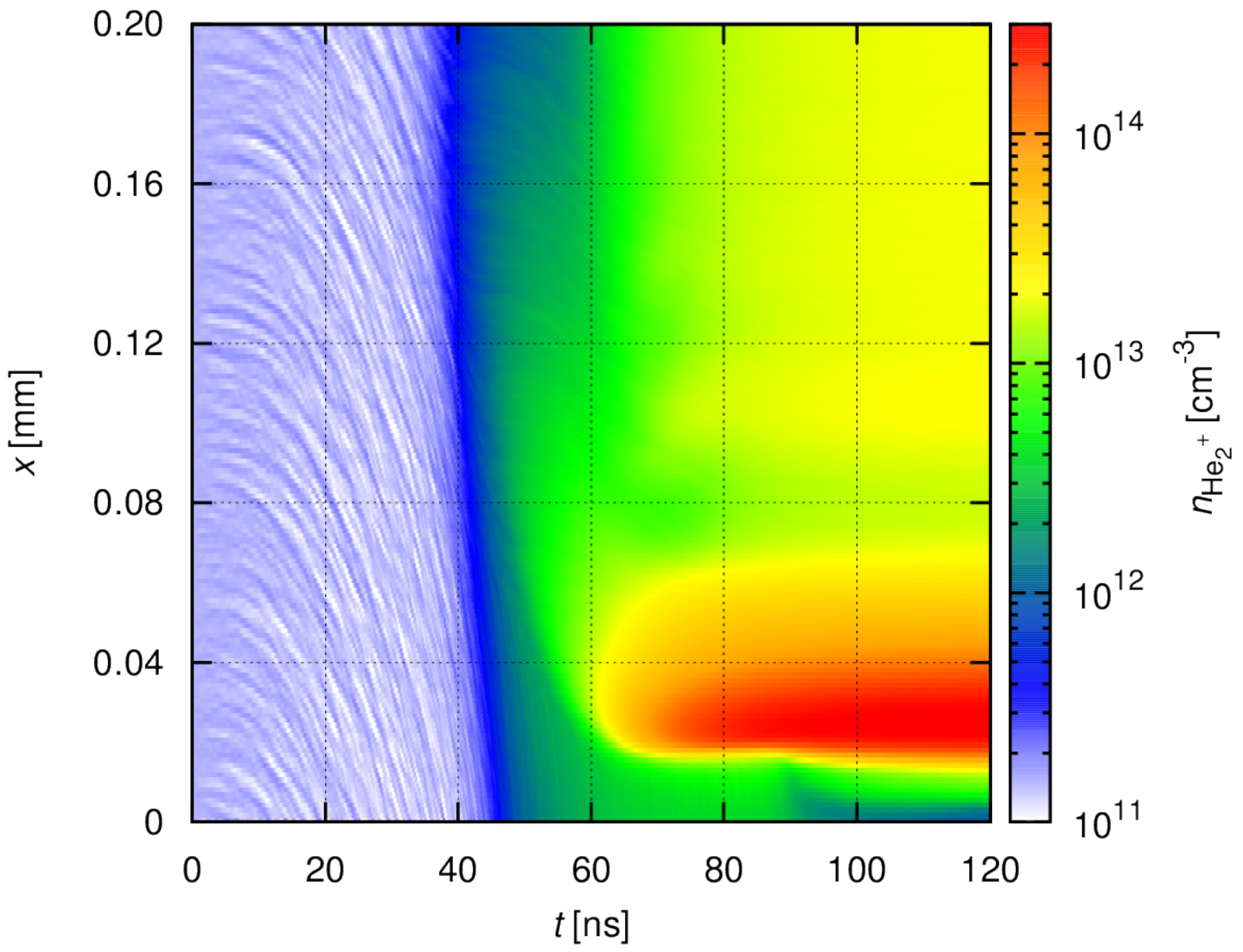}\\
\small (c) \includegraphics[width =0.5\textwidth]{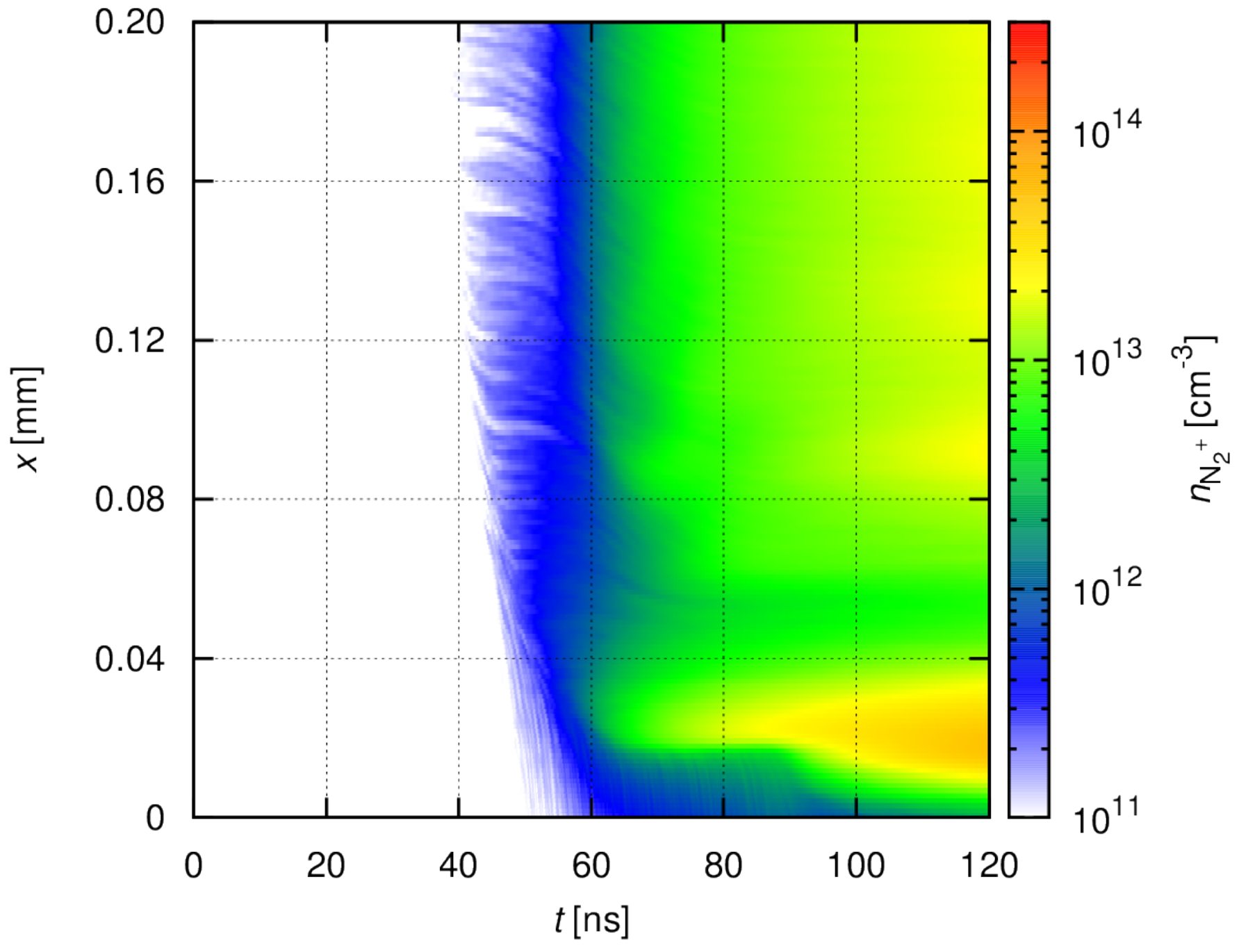}
\caption{The spatio-temporal distribution of densities of (a) He$^+$ ions, (b) He$_2^+$ ions, and (c) N$_2^+$ ions for the base conditions (He + 0.1\% N$_2$, at $p$ = 1 bar,  $L$ = 1 mm, $V_0$ = 1000 V, $\tau_{\rm p}$ = 30 ns, and $n_0 = 1.5 \times 10^{11}$ cm$^{-3}$, with VUV photons considered). These figures show only the region adjacent to the cathode, the density change from the $x$ = 0.2 mm position towards the anode is insignificant.}
\label{fig:idensity}
\end{center}
\end{figure}

The spatio-temporal evolution of the electron density is shown in figure \ref{fig:edensity}. At early times, below $\sim$35 ns, we observe that the initial electrons are accelerated towards the anode (located at $x$ = 1 mm) by the increasing voltage. At high voltage, clearly visible electron avalanches starting from the cathode and from the gas phase (due to the creation of free electrons in Penning and electron-impact ionization processes) develop and lead to the breakdown of the gas. During this process the electron density increases by orders of magnitude and the electron-rich region rapidly approaches the cathode. Beyond $\sim$\,55-60\,ns the electron density is depleted in a very narrow cathode sheath only, then exhibits its prominent peak in the negative glow at $x \approx 0.02$ mm, where electrons are trapped in the reversed-field region (cf. figure~\ref{fig:field-disp}(a)).  Figure \ref{fig:edensity}  also reveals the signature of the Faraday dark space and the formation of a nearly homogeneous positive column region, with striations attached to the Faraday dark space (in the domain between $x \cong$ 0.1 and 0.2 mm, around 60-80 ns, before the termination of the voltage pulse). Following the termination of the voltage pulse the cathode sheath disappears within a short time (order of few ns). The further development of the electron density beyond the termination of the excitation pulse is governed by the ambipolar field that forms due to the presence of different (positive) ionic species. We note that this decay is slow as the Penning ionization process acts as a source of electrons, well beyond the termination of the voltage pulse, as well.

\begin{figure}[ht ]
\begin{center}
\small (a) \includegraphics[width =0.5\textwidth]{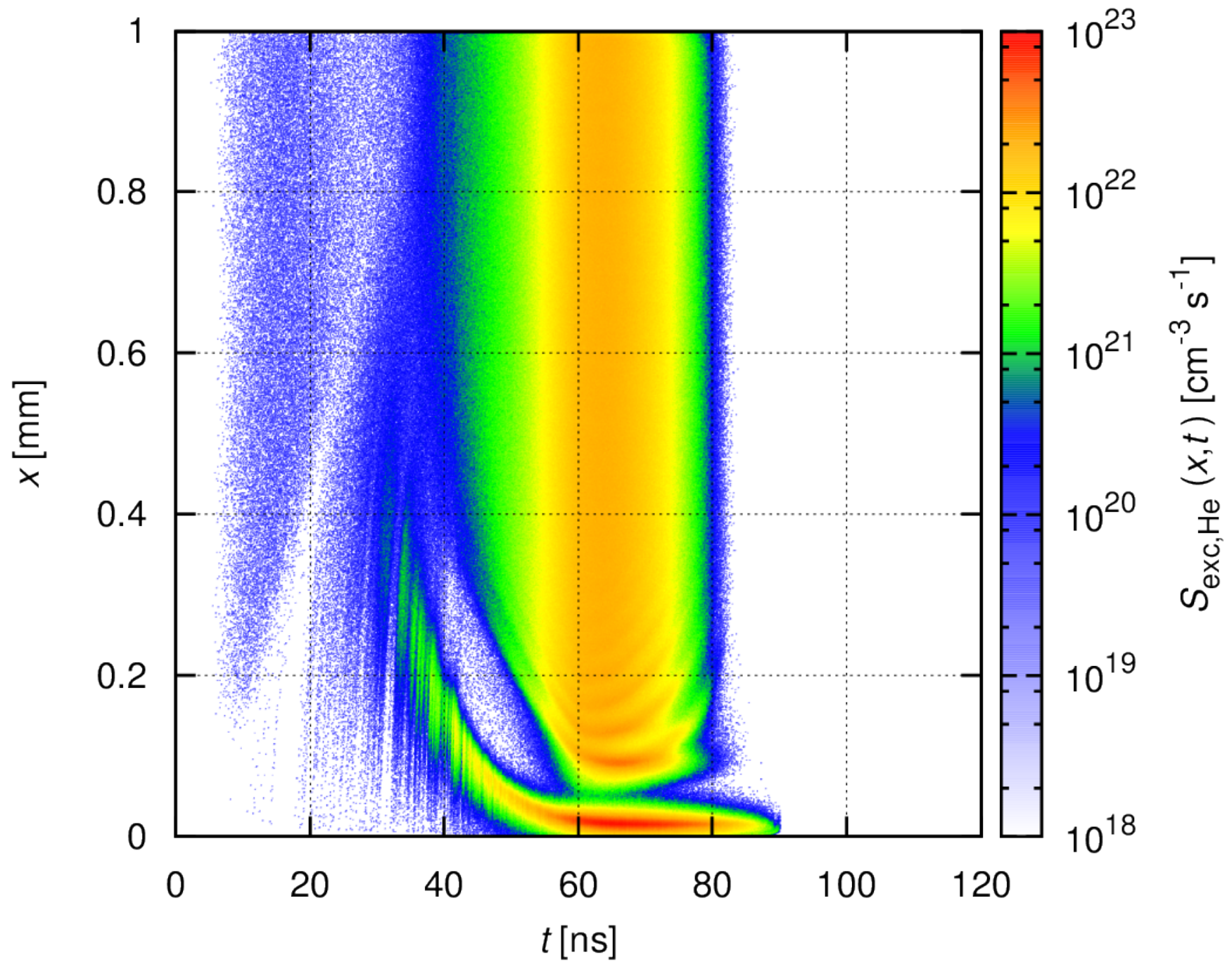}\\
\small (b) \includegraphics[width =0.5\textwidth]{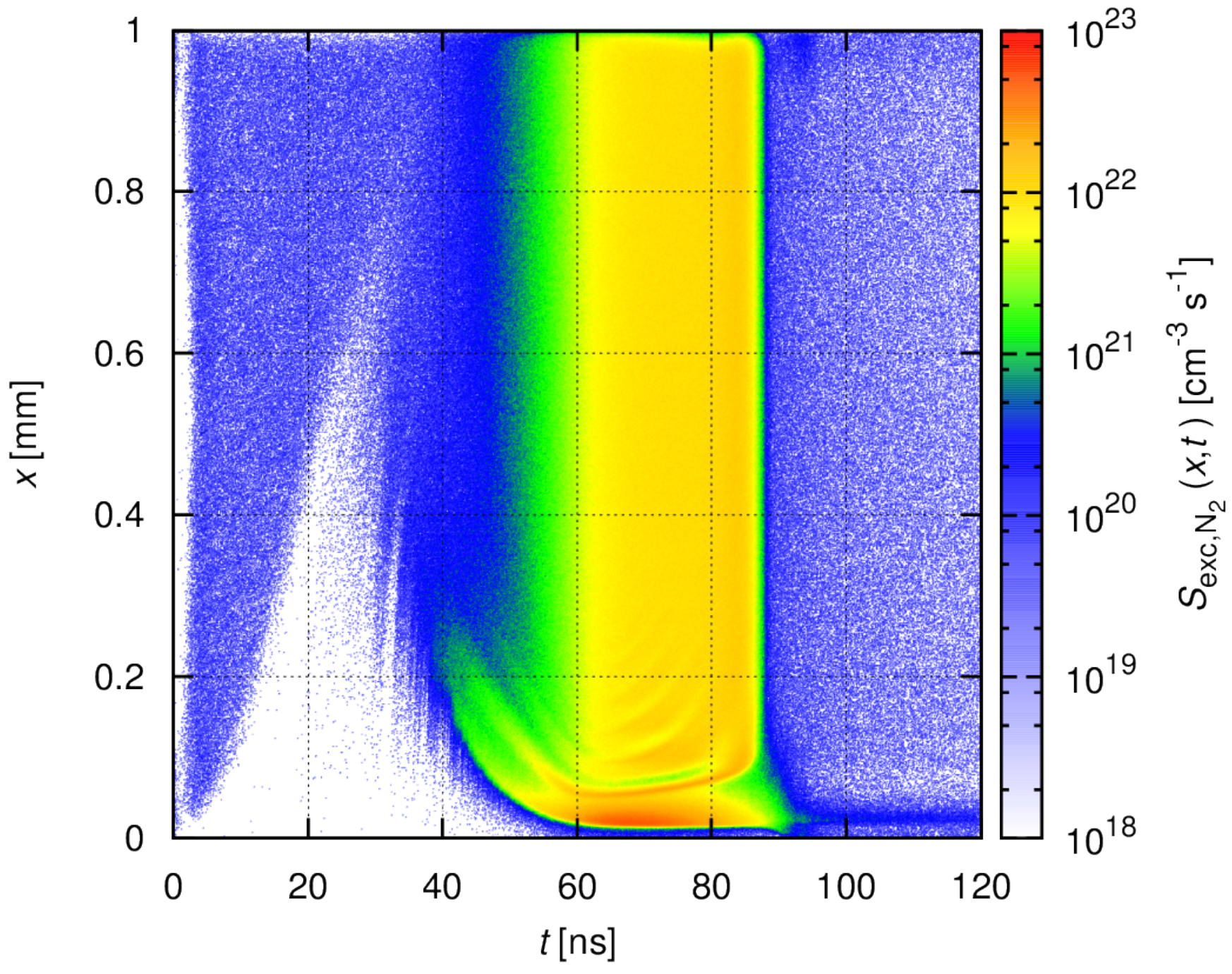}
\caption{The spatio-temporal distribution of electron impact excitation rates of (a) He atoms and (b) N$_2$ molecules (including all electron-impact excitation channels) for the base conditions (He + 0.1\% N$_2$, at $p$ = 1 bar,  $L$ = 1 mm, $V_0$ = 1000 V, $\tau_{\rm p}$ = 30 ns, and $n_0 = 1.5 \times 10^{11}$ cm$^{-3}$, with VUV photons considered).}
\label{fig:excrates}
\end{center}
\end{figure}  

The ion composition of the plasma is revealed in figures \ref{fig:idensity}(a)-(c), which show the densities of He$^+$, He$_2^+$, and N$_2^+$ ions, respectively in the spatial domain 0 mm $\leq x \leq$ 0.2 mm. (Beyond $x=0.2$ mm the densities are nearly the same within the rest of the gap as at $x=0.2$ mm.) The highest He$^+$ ion density is reached between 60 and 90 ns, around the spatial position $x \cong 0.025$ mm (in the negative glow). A decay of density is observed later on due to the lack of further source due to electron impact and due to the conversion of He$^+$ ions to He$_2^+$ molecular ions.  The density of the latter is also highest in the negative glow region (see figure \ref{fig:idensity}(b)). The peak densities are in the range of $2\,\times\,10^{14}$\,cm$^{-3}$. The N$_2^+$ ions have about an order of magnitude lower density compared to the helium ionic species. This density is, however, still remarkable, considering the low, 0.1\% concentration of N$_2$ in the background gas. It is noted that the atomic helium ions form the major part of the positive space charge in cathode sheath region, which is explained by the fact that their cross section (including the resonant charge transfer reaction) is higher than the Langevin cross sections of the other ionic species.

The excitation rates of He atoms and N$_2$ molecules by electron impact are displayed in figure \ref{fig:excrates}. At early times, $t \leq$\,30\,ns we observe excitation due to the acceleration of the "seed" electrons towards the anode. A prominent feature of the excitation rate in figure \ref{fig:excrates}(a) marks the edge of the developing negative glow. For both He and N$_2$, the most intense excitation is observed in the glow region, around $x \approx 0.02$\,mm, at times between 60 ns and 80 ns. Within this time domain, following a spatial decay of the excitation rate (corresponding to the Faraday dark space), the column region is seen to appear, with a striated structure attached to the Faraday dark space at times when electron emission from the cathode is appreciable. The periodicity of the striations changes as a function of the varying voltage amplitude. Excitation of N$_2$ persists for a longer time, due to the lower excitation energies of N$_2$ molecules, compared to those of He. Excitation of N$_2$ also appears in the afterglow domain. This excitation corresponds to the rotational excitation. 

\begin{figure}[ht ]
\begin{center}
\includegraphics[width =0.5\textwidth]{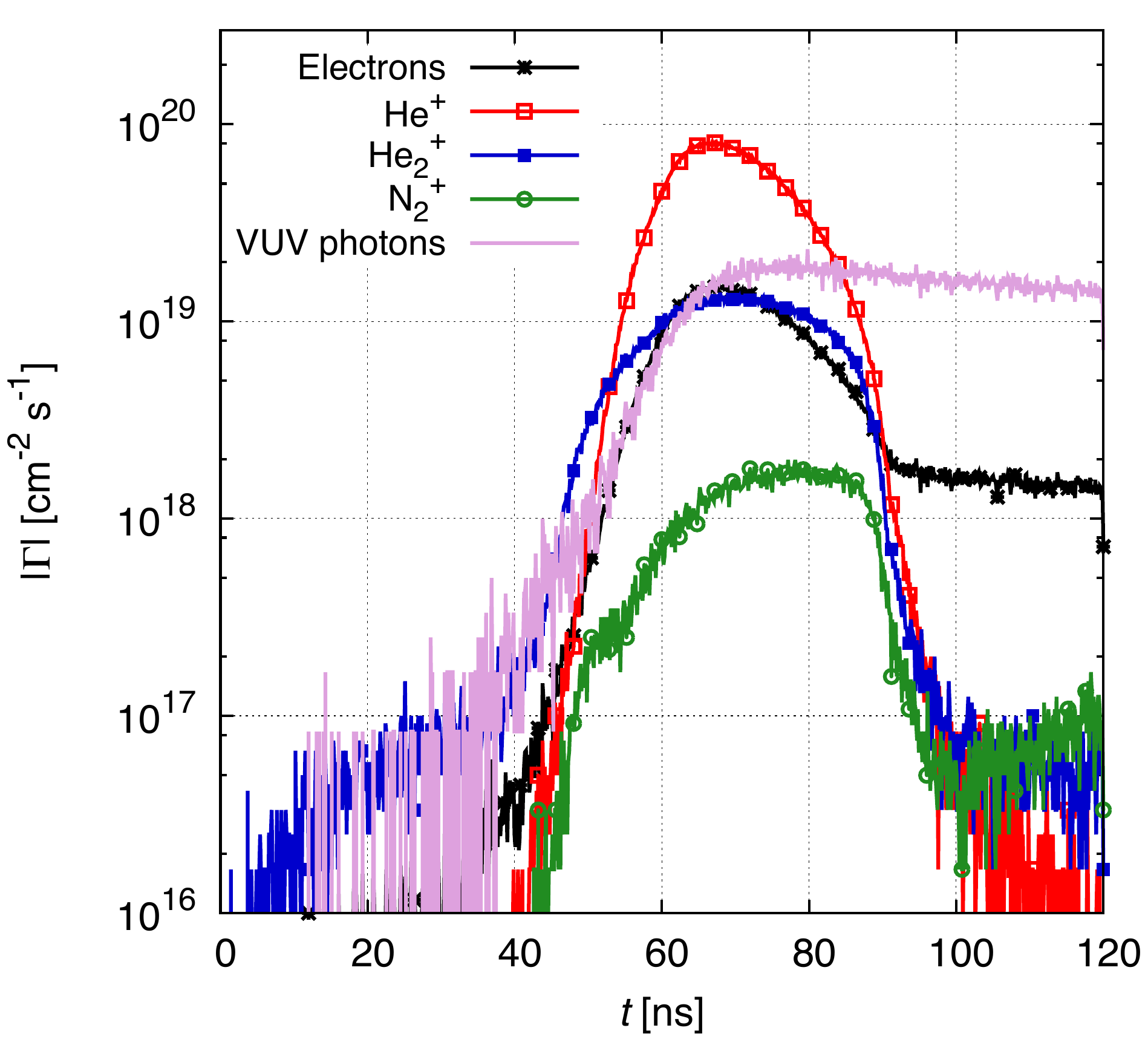}\\
\includegraphics[width =0.5\textwidth]{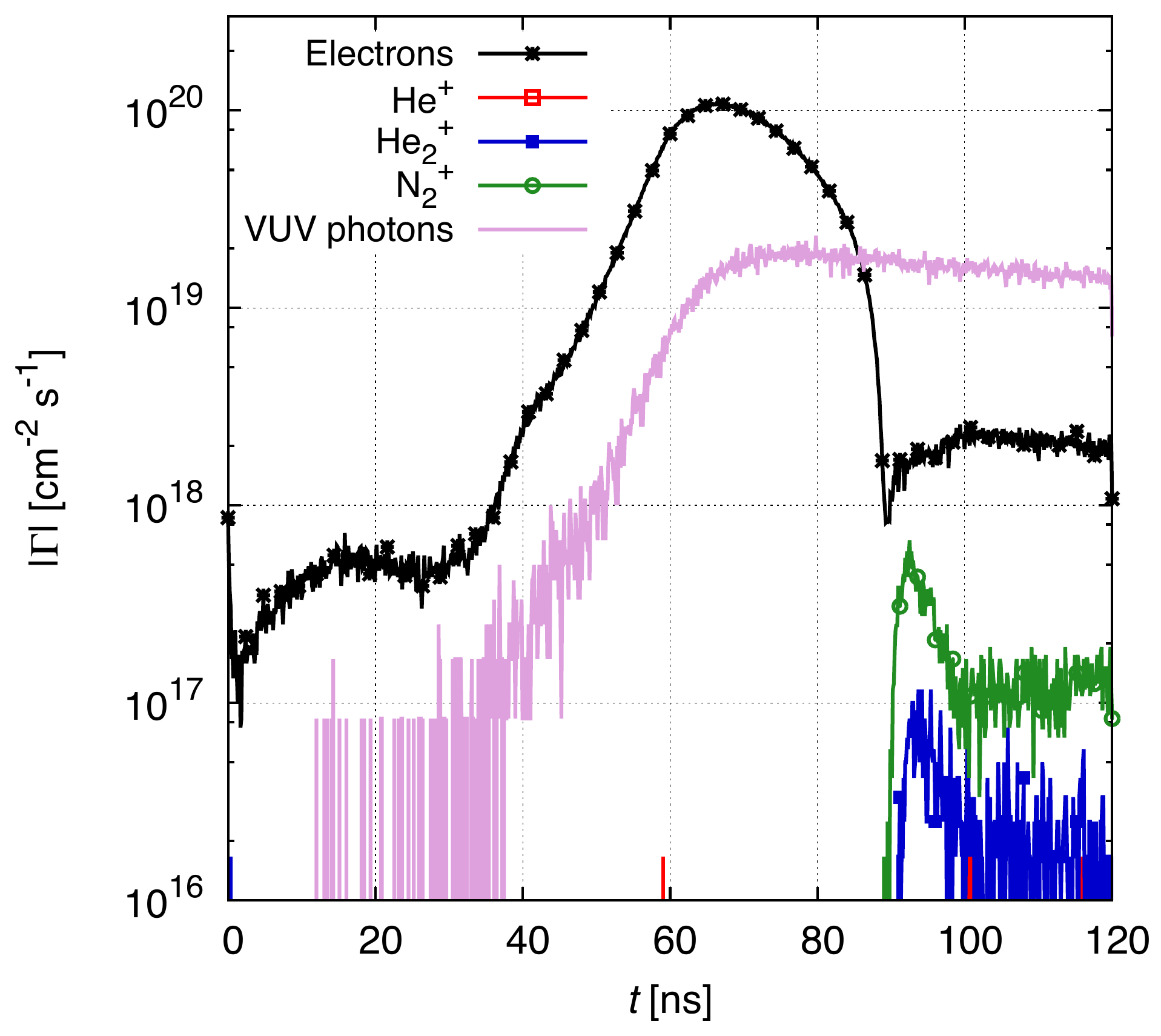}
\caption{Time dependence of the fluxes of different charged species and VUV resonance photons at (a) the cathode and (b) the anode, at the base conditions (He + 0.1\% N$_2$, at $p$ = 1 bar,  $L$ = 1 mm, $V_0$ = 1000 V, $\tau_{\rm p}$ = 30 ns, and $n_0 = 1.5 \times 10^{11}$ cm$^{-3}$). The data for the ions and photons represent fluxes {\it reaching} the electrodes. For electrons the net flux is shown that includes secondary emission; (a) shows the electron flux {\it leaving} the cathode, while (b) shows the electron flux {\it reaching} the anode.}
\label{fig:fluxes}
\end{center}
\end{figure}

The fluxes of the charged particles and VUV photons at the electrodes are shown in figure \ref{fig:fluxes}; panel (a) shows the fluxes at the cathode and panel (b) at the anode. The data for the ions and photons represent fluxes {\it reaching} the electrodes. For electrons the net flux is shown that includes secondary emission; for the cathode the electron flux {\it leaving} the electrode, while for the anode the flux {\it reaching} the electrode is shown.

Among the species considered, some of the "seed" He$_2^+$ ions and VUV photons arrive earliest at the cathode. The electrons emitted by these species have primary importance in the development of the breakdown process under the conditions of the temporally rapidly increasing voltage at $t \leq$ 30\,ns. During the plateau phase of the voltage pulse, atomic helium ions take over dominating the flux to the cathode, while beyond $t=90$\,ns the charged particle fluxes rapidly drop and only the photon flux remains high, in the 10$^{19}$\,cm$^{-2}$s$^{-1}$ range. The photon flux at the anode is very similar in both magnitude and time-dependence to that at the cathode. During the voltage pulse the flux at the anode is (trivially) dominated by electrons. Beyond $t=$\,90\,ns their net flux drops to level of $\approx 2 \times 10^{18}$ cm$^{-2}$s$^{-1}$. At the same time fluxes of ions (mainly those of N$_2^+$ and He$_2^+$) appear at a level of $10^{16} - 10^{17}$ cm$^{-2}$s$^{-1}$. At a later stage (well beyond the time limit of the simulations) the positive ion and electron fluxes are expected to balance each other at both electrodes under the conditions of the ambipolar transport.

\begin{figure}[ht ]
\begin{center}
\includegraphics[width =0.5\textwidth]{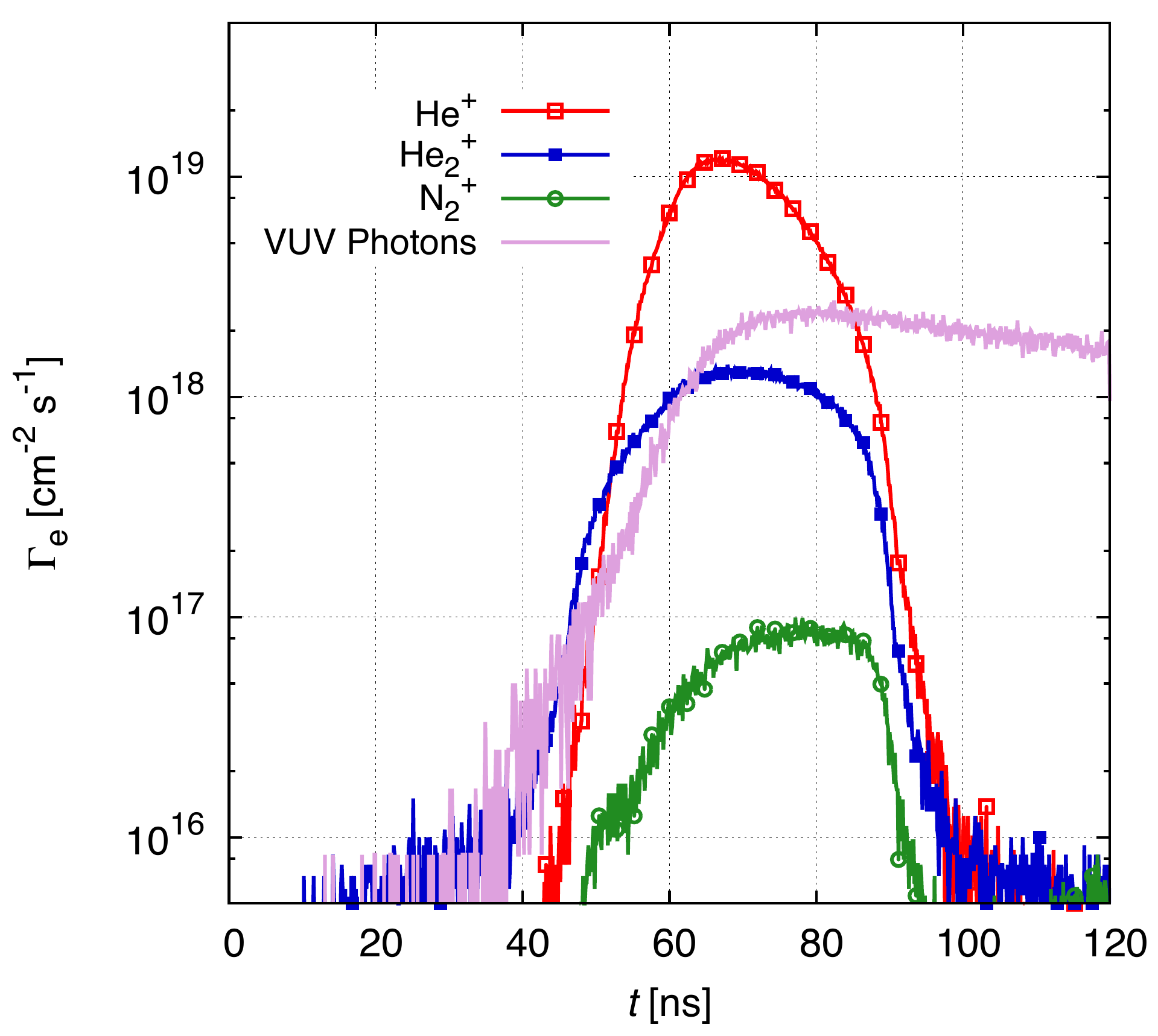}
\caption{Time dependence of the flux of electrons emitted from the cathode due to the impact of different species, at the base conditions (He + 0.1\% N$_2$, at $p$ = 1 bar,  $L$ = 1 mm, $V_0$ = 1000 V, $\tau_{\rm p}$ = 30 ns, and $n_0 = 1.5 \times 10^{11}$ cm$^{-3}$).}
\label{fig:efluxes}
\end{center}
\end{figure}

The flux of electrons emitted from the cathode due to the different species (ions and VUV photons) is displayed in figure \ref{fig:efluxes}. During the breakdown phase He$_2^+$ ions (the only ionic species considered at the initialization of the simulations) and photons contribute to secondary electron emission. Subsequently, beyond $t \approx$ 50 ns, He$^+$ ions take over and dominate the production of secondary electrons, while the share of He$_2^+$ and N$_2^+$ ions is, respectively, at the 10\% and 1\% level. Due to the trapping of the resonance photons they release electrons from the cathode even well after the termination of the voltage pulse. 

\begin{figure}[h!]
\begin{center}
\small \small (a) \includegraphics[width =0.5\textwidth]{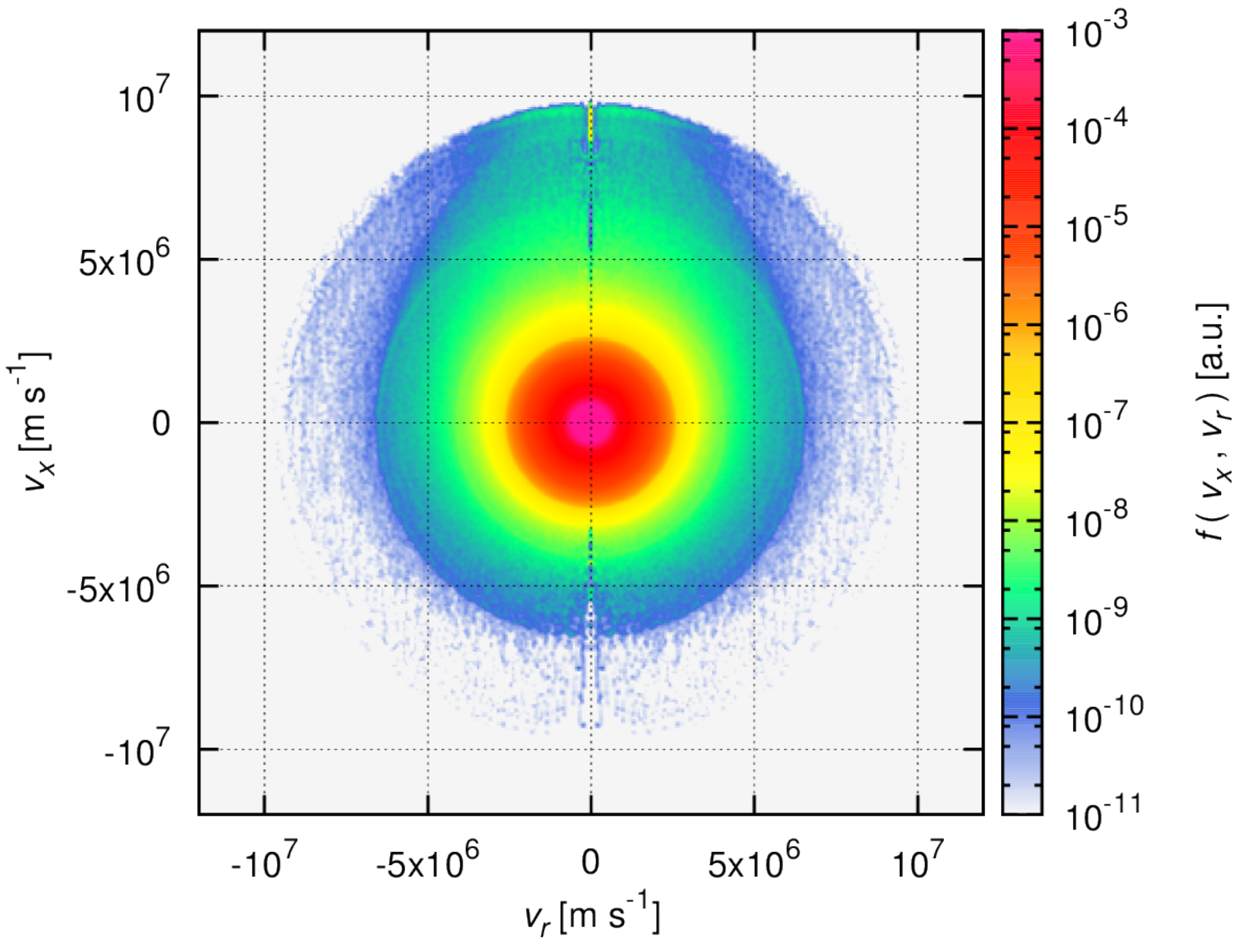}\\
\small \small (b) \includegraphics[width =0.5\textwidth]{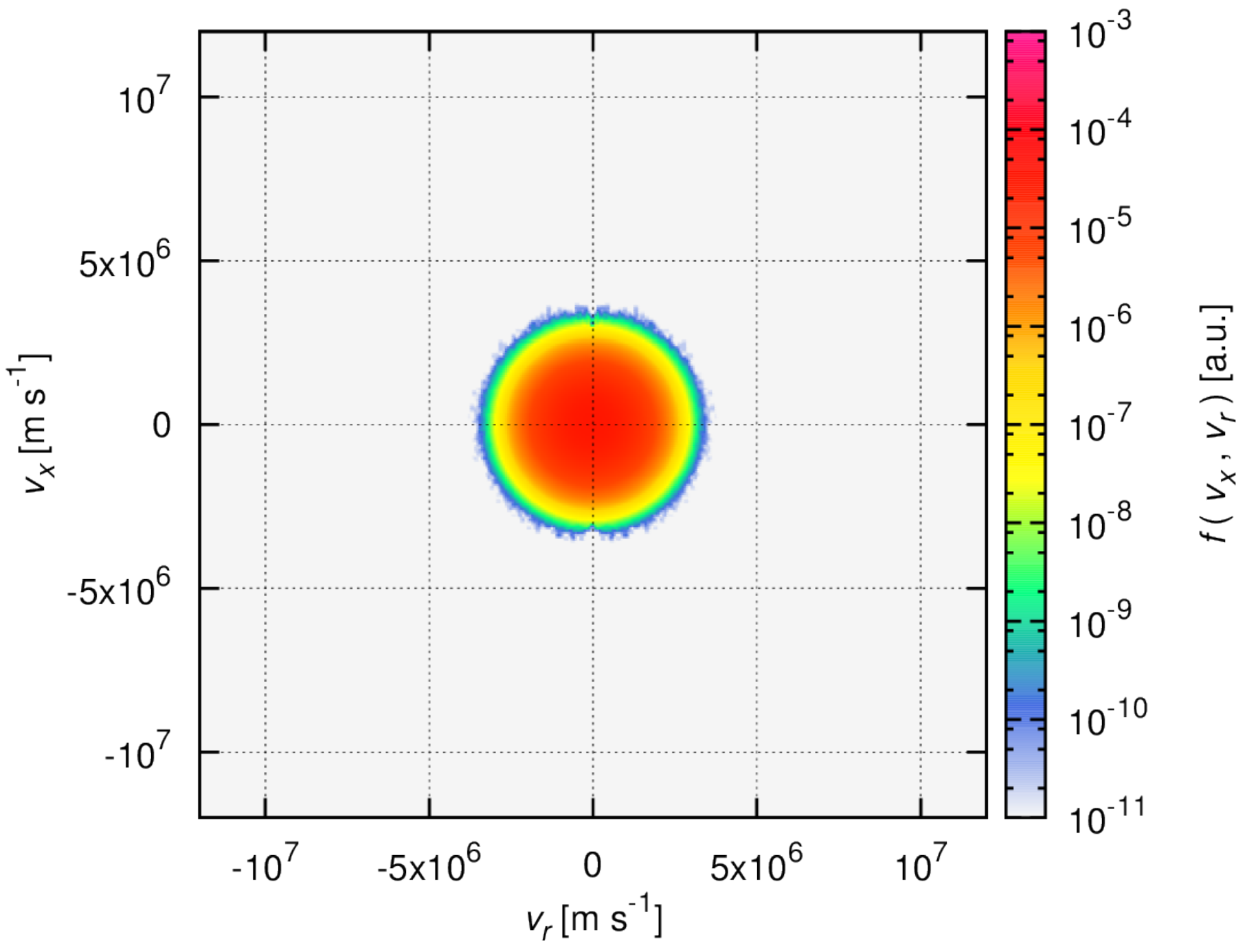}\\
\caption{Velocity distribution functions of the electrons at 70 ns, at the spatial positions (a) $x=0.02$ mm and (b) $x=0.60$ mm. The temporal and spatial window for the data collection are, respectively, $\pm 1$ ns and $\pm$ 0.01 mm. Position (a) is near the sheath--glow boundary, position (b) belongs to the positive column. Discharge conditions: He + 0.1\% N$_2$, $p$ = 1 bar,  $L$ = 1 mm, $V_0$ = 1000 V, $\tau_{\rm p}$ = 30 ns,  $n_0 = 1.5 \times 10^{11}$ cm$^{-3}$, VUV photons considered.}
\label{fig:vdf}
\end{center}
\end{figure}

\begin{figure}[h!]
\begin{center}
\includegraphics[width =0.5\textwidth]{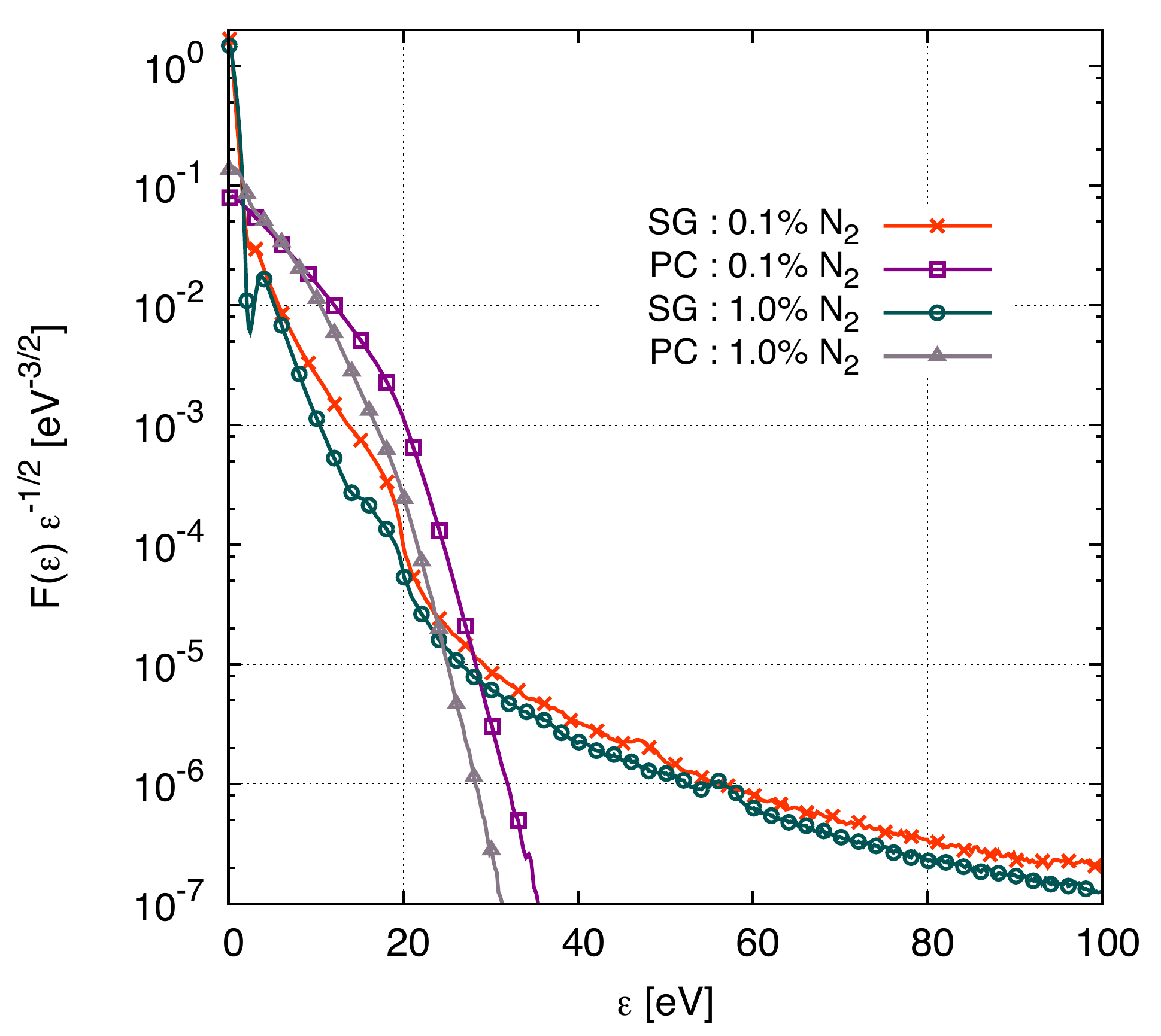}
\caption{Electron energy distribution functions at 70 ns, at the spatial positions (a) $x=0.02$ mm (corresponding to sheath -- negative glow boundary, "SG") and (b) $x=0.60$ mm (which belongs to the positive column, "PC"). The temporal and spatial window for the data collection are, respectively, $\pm 1$ ns and $\pm$ 0.01 mm. Results are shown for 0.1\% and 1.0\% N$_2$ concentrations, at $p$ = 1 bar,  $L$ = 1 mm, $V_0$ = 1000 V, $\tau_{\rm p}$ = 30 ns, and $n_0 = 1.5 \times 10^{11}$ cm$^{-3}$ (with VUV photons considered).}
\label{fig:edf}
\end{center}
\end{figure}

\begin{figure}[h!]
\begin{center}
~~~~~~~~~ \small (a)\includegraphics[width =0.52\textwidth]{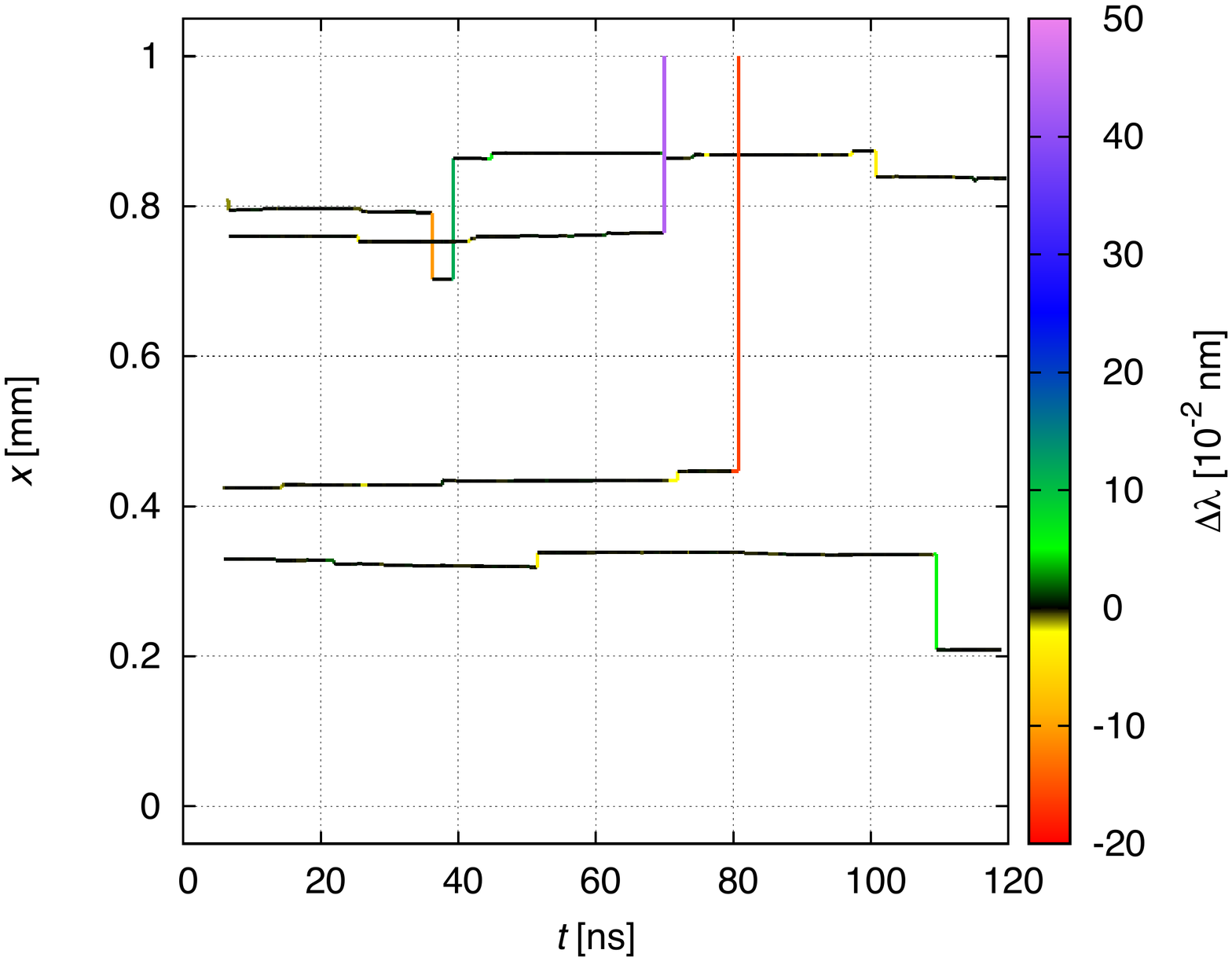}\\
~~~~~~~~~~~ \small (b)\includegraphics[width =0.52\textwidth]{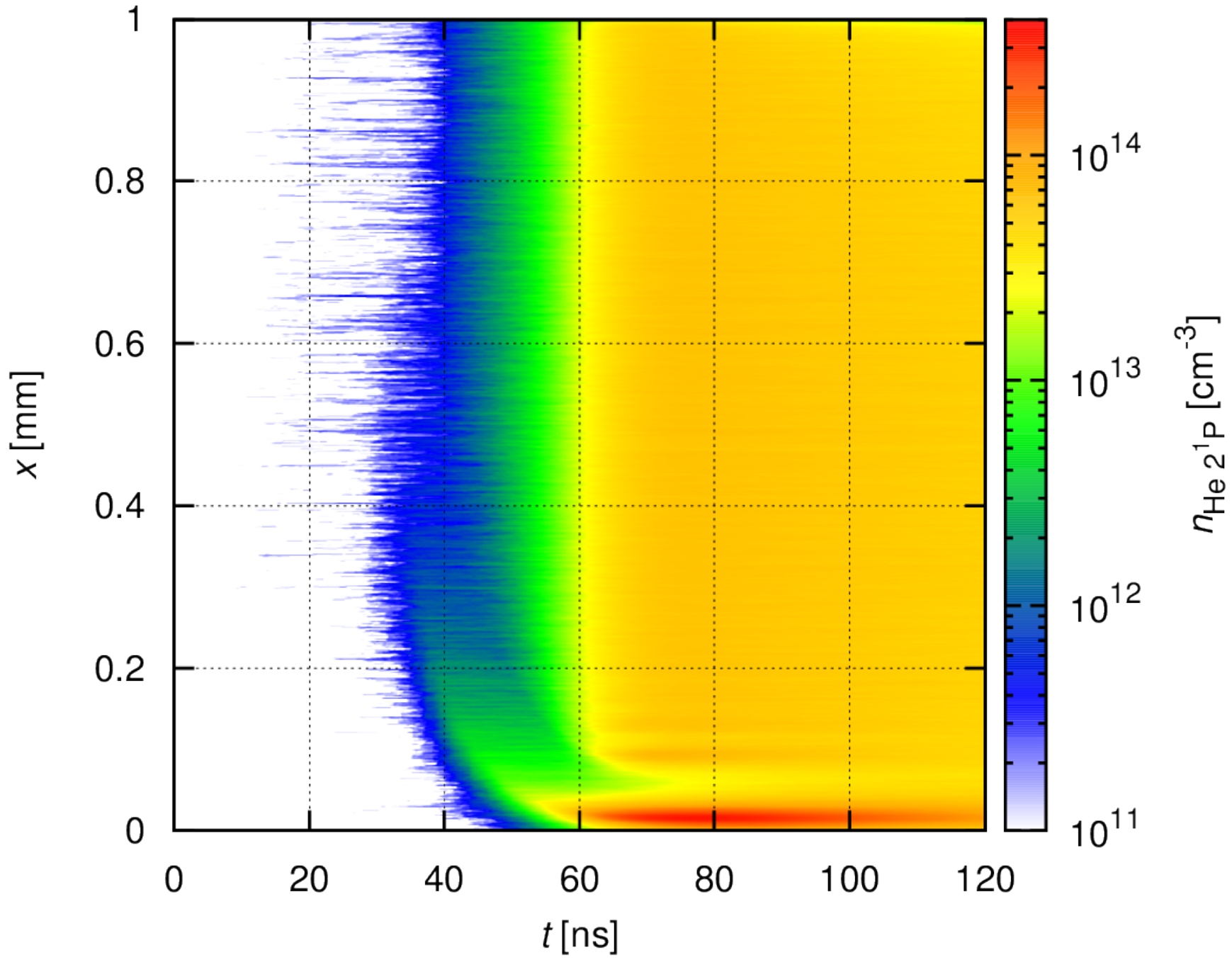}\\
\small (c) \includegraphics[width =0.44\textwidth]{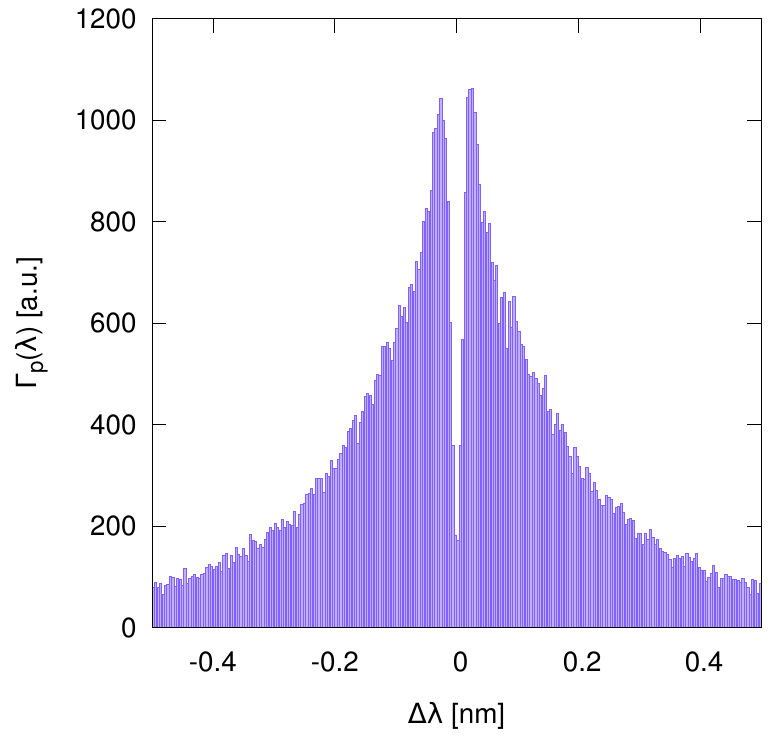}
\caption{(a) The path of four photons from the first emission event to the absorption at an electrode or to the end of the simulation (120 ns). The actual deviation of the wavelength from the line center ($\Delta \lambda$) is color coded. (b) Spatio-temporal distribution of the He 2$^1$P excited state density and (c) time-averaged distribution of the wavelengths of VUV photons (photon flux in arbitrary units) reaching the cathode, around the central wavelength $\lambda_0 = 58.4334 $ nm, at the base conditions. (He + 0.1\% N$_2$, $p$ = 1 bar,  $L$ = 1 mm, $V_0$ = 1000 V, $\tau_{\rm p}$ = 30 ns, $n_0 = 1.5 \times 10^{11}$ cm$^{-3}$.)}
\label{fig:photons}
\end{center}
\end{figure}

The velocity distribution functions (VDF) of the electrons are also analyzed. Due to the symmetry of the system, the "general" VDF $f({\bf v},{\bf r},t)$ reduces to $f(v_x,v_r,x,t)$. Two examples of this function are illustrated in figures \ref{fig:vdf}(a) and (b). The first panel shows the VDF around $(x,t)$ = (0.02 mm, 70 ns), while the second corresponds to (0.69 mm, 70 ns) coordinates. Data for the VDFs have been collected within spatial and temporal domains with finite width, $(\Delta x)_{\rm VDF} = \pm$ 0.01 mm and $(\Delta t)_{\rm VDF} = \pm$ 1 ns. Figure \ref{fig:vdf}(a) shows a very clear directional anisotropy of the distribution function, as well as the presence of a significant number of high energy electrons. Even beam electrons (propagating away from the cathode), which reached the data collection region from the cathode without inelastic energy losses, are seen at $v_x \approx 0.95 \times 10^7$ m s$^{-1}$ and $v_r \approx 0$. The corresponding electron energy distribution function also shows the presence of a remarkable high-energy tail, see the "SG : 0.1\% N$_2$" curve of figure \ref{fig:edf}. These observations show a large degree of similarity with the case of the cathode region of low-pressure discharges \cite{Boeuf_Marode,Donko2011} and point out the need for kinetic simulations even at high pressures. Further away from the cathode, in the column region (see figure \ref{fig:vdf}(b)) the VDF appears to be isotropic and the EEDF ("PC : 0.1\% N$_2$" curve) exhibits a cutoff that is characteristic at the conditions of a low electric field. 

\vspace{0.5cm}

As already mentioned in connection with figure~\ref{fig:currents1}, at the conditions considered, it is the peculiarity of photon transport that during the repeating emission -- reabsorption events the time spent by the atoms in the excited state is much longer than the free "flights" of the photons. This way the number of excited atoms is much higher in the simulation than the number of photons. Further simulation results relevant to the transport of photons are revealed in figure~\ref{fig:photons}. The pathways of few individual photons can be followed in figure~\ref{fig:photons}(a). The four trajectories originate from four separate simulations, and in each of these the first photon created was selected for this analysis. The photons are created at $\approx$ 5 ns and two of them reached one electrode during the time frame of the simulation. The trajectory segments are color coded according to wavelength of the photon that changes after each absorption -- re-emission event (as described in section 2.3). The wavelength at the line center (at $\Delta \lambda = 0$ that corresponds to $\lambda = \lambda_0$) is shown by black color. At a wavelength very close to $\lambda_0$ the photons have very short free path due to the property of the photoabsorption cross section that it sharply peaks at $\lambda_0$. Accordingly, the $x(t)$ trajectories are (nearly) horizontal, and consist of numerous small segments (which are not resolved in this plot). As most of the emission events generate photons "near" $\lambda_0$,  the photons are strongly trapped inside the plasma. Whenever the photon is radiated further away from the line center (i.e. at the wings of the spectral line), longer free paths are possible, which show up in nearly vertical segments (representing a large displacement in the $x$ direction during one flight segment) in figure~\ref{fig:photons}(a). These segments show up in different colors according to the actual wavelength deviation $\Delta \lambda$. We note that, for these four sample cases the number of emission and reabsorption events was between 90 and 217, indicating the high importance of these processes in the photon propagation.

Figure~\ref{fig:photons}(b) gives the density of the excited resonant state. This spatio-temporal distribution resembles, both in shape and magnitude, that of the charged particles. The highest density is found beyond $t \approx 60$ ns, in the vicinity of the cathode. Figure~\ref{fig:photons}(c) shows the spectral distribution of the wavelength of the VUV photons reaching the cathode. Due to the very high photoabsorption cross section the radiation near the line center is absorbed within a very short distance, thus radiation in this wavelength range practically cannot leave the plasma (cf.  figure~\ref{fig:photons}(a)), unless the excited atom radiates very near the boundary of the domain examined, i.e. near the electrode. The dip that forms at the line center is typical for high-pressure conditions. At low pressure, when the photons have longer free path, this dip gradually disappears (see e.g. \cite{Fierro}).

\subsection{Parameter variation}

\label{sec:res2}

Due to the high number of parameters (pressure, gap length, voltage, pulse duration, etc.) we can present in this article only a small subset of a systematic parametric investigation of the model system. Below, we shall illustrate the effects of the initial density and the nitrogen concentration, and give a few examples of the current pulse development for selected values of the voltage amplitude and the pulse length.

\begin{figure}[h]
\begin{center}
\small (a) \includegraphics[width =0.44\textwidth]{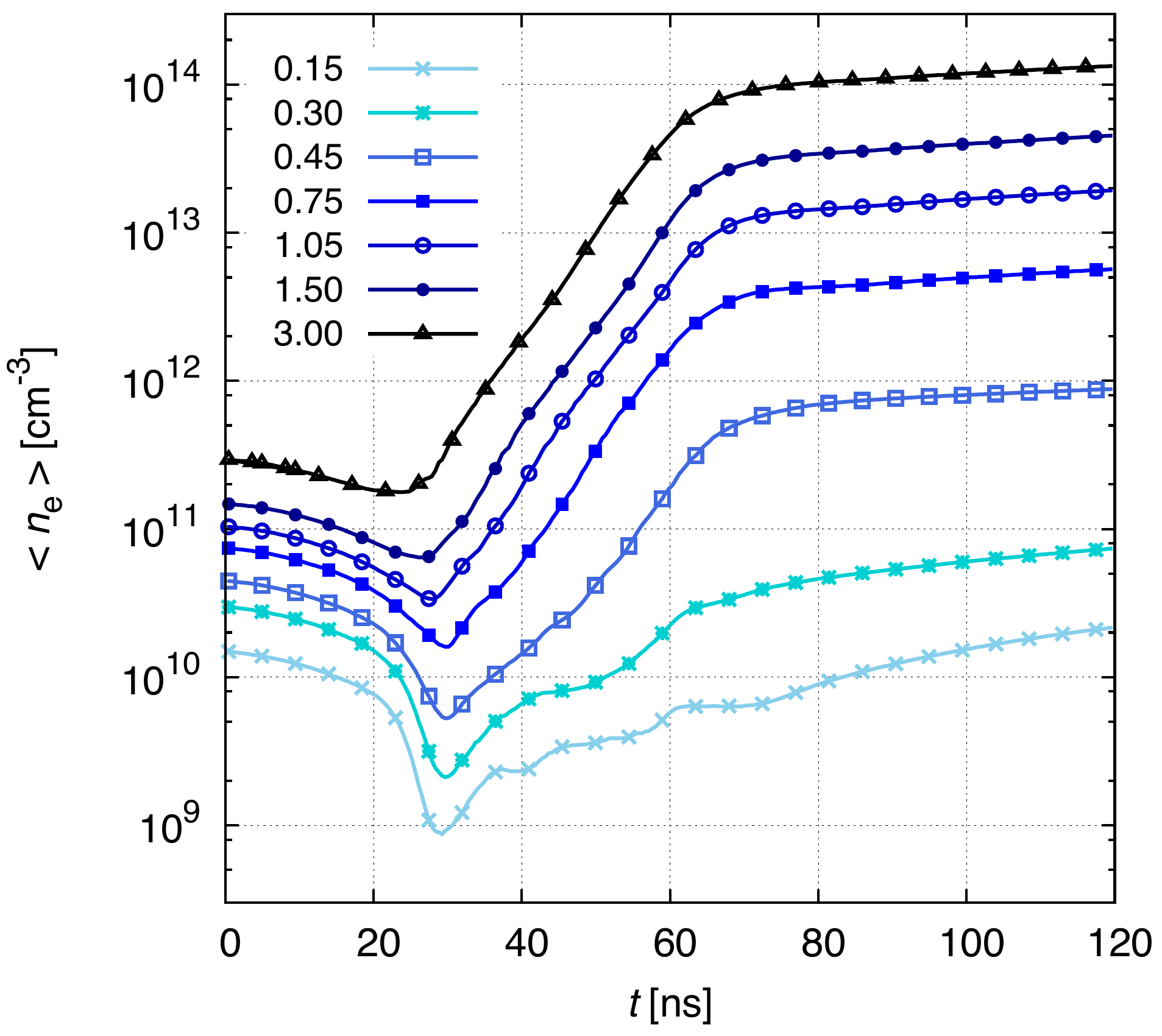}\\
\small (b) \includegraphics[width =0.44\textwidth]{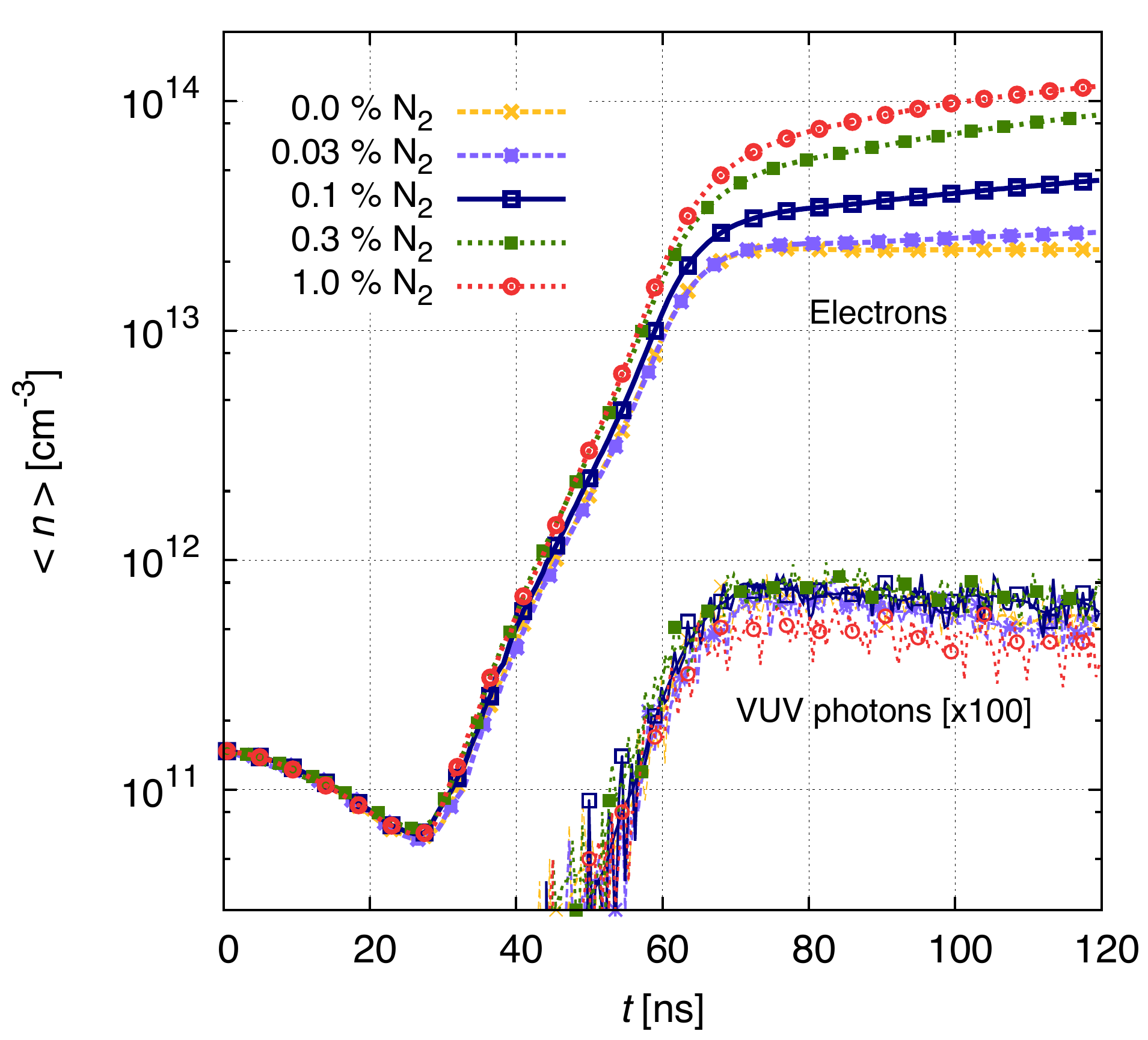}
\caption{Time dependence of the spatially averaged electron density at various values of the initial electron density (a) and at different values of the N$_2$ concentration in the background gas (b). For (a) the N$_2$ content is 0.1\% and the legend gives the initial electron density $n_0$ in units of 10$^{11}$ cm$^{-3}$. For (b) the initial electron density is 1.5 $\times$ 10$^{11}$ cm$^{-3}$. (Note that the curves for the VUV photons are 100$\times$ multiplied.) Other conditions: $p$ = 1 bar, $L$ = 1 mm, $V_0$ = 1000 V, $\tau_{\rm p}$ = 30 ns for all cases.}
\label{fig:initdens}
\end{center}
\end{figure}

The results of a study of the effect of the initial value of the electron density on its time dependence are shown in figure \ref{fig:initdens}(a). The initial density was varied between $n_{\rm 0} = 10^{10}$\,\,cm$^{-3}$ and $2 \times 10^{11}$ cm$^{-3}$. At the lowest $n_{\rm 0}$ the mean electron density at the end of the simulation is approximately the same as the initial density and the peak of the current pulse is as low as 0.03 A. The $n_{\rm 0} = 1.5 \times 10^{10}$\,\,cm$^{-3}$ value corresponds to the base case; here the mean electron density increases by a factor $\sim 200$ by the end of the voltage pulse. These results indicate that the initial density has a strong influence of the temporal development of the electron density, and consequently on the peak value of the current pulse as well (not shown). Figure \ref{fig:initdens}(b) displays the mean electron and photon density at selected values of N$_2$ admixture concentrations. A higher N$_2$ content is found to increase the electron density, which is a consequence of the charged particle production via the electron-impact and Penning ionization processes. The density (or number) of photons, on the other hand, slightly decreases with increasing N$_2$ content, which can be explained by the increasing excitation rate of N$_2$ and a consequent depletion of the density of high energy electrons, which, in turn, leads to a lower excitation rate of helium atoms. This effect is confirmed in figure \ref{fig:edf} that shows a significant decrease of the number of high energy electrons as the nitrogen concentration increases (compare the "0.1\% N$_2$" and "1.0\% N$_2$" curves).

\clearpage

\begin{figure}[ht ]
\begin{center}
\includegraphics[width =0.5\textwidth]{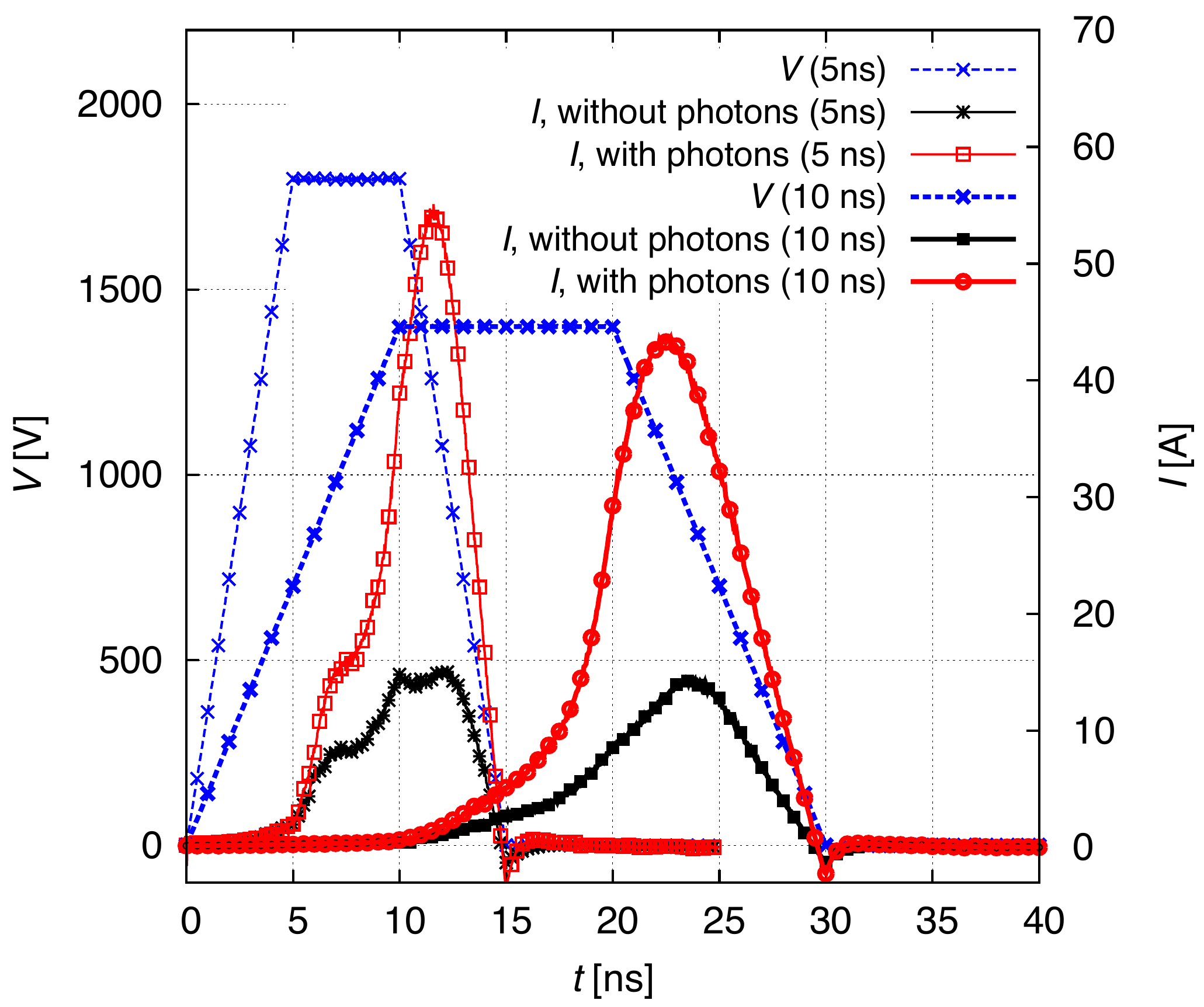}
\caption{Excitation voltage waveforms (blue dashed lines) and discharge current  obtained in the simulations including (red solid lines) and excluding (black solid lines) photons, for He + 0.1\% N$_2$, at $p$ = 1 bar and $L$ = 1 mm, for pairs of peak voltage and plateau duration values of $V_0$ = 1800 V /$\tau_{\rm p}$ = 5 ns (thin lines) and $V_0$ = 1400 V / $\tau_{\rm p}$ = 10 ns (thick lines).}
\label{fig:par-currents}
\end{center}
\end{figure}

Simulations have also been conducted at various voltage pulse durations ($\tau_{\rm p}$) and amplitudes (plateau values, $V_0$). Two examples are shown in figure~\ref{fig:par-currents}, which shows the cases of the $V_0$ = 1800 V / $\tau_{\rm p}$ = 5 ns and $V_0$ = 1400 V / $\tau_{\rm p}$ = 10 ns parameter pairs. The inclusion of VUV photons in these two cases also confirms their important role, the current pulse amplitudes are about 3-4 times higher when electron emission from the cathode due to the photoeffect is taken into account. While a parametric investigation covering a wide parameter domain is beyond the scope of the current study, we contemplate that a similar qualitative behavior regarding the effect of VUV photons is present across parameter ranges.   

\section{Summary}

We have investigated the effect of VUV resonance radiation, acting via the photoemission of electrons from the electrodes, on the characteristics of atmospheric-pressure microdischarges excited by $\sim$ kV-amplitude voltage pulses of 15-90 ns duration. Most of the computations have been conducted for He gas with a small, 0.1\% admixture of N$_2$, but other concentrations between 0\% and 1\% N$_2$ admixture were also considered. The study has been based on the Particle-in-Cell / Monte Carlo Collisions approach, into which the transport of VUV photons was also incorporated in a particle-based manner. The timing of the simulation was carefully set to comply with the relevant stability criteria and ensuring maximum possible efficiency.

A detailed analysis has been presented for a base case of a voltage pulse of $\tau_{\rm p} =$30\,ns plateau duration and $V_0$ = 1000 V amplitude, for He + 0.1\% N$_2$ gas mixture, at $p$ = 1 bar pressure and $L$ = 1 mm electrode gap. The simulations provided insight into the development of the discharge via the computation and the analysis of the spatio-temporal distributions of the densities of the different charged species (electrons, He$^+$ ions, He$_2^+$ ions, N$_2^+$ ions), He metastable atoms,  He 2$^1$P resonant state atoms, and VUV photons. The fluxes of the charged species and VUV photons revealed the importance of the photoemission process from the cathode. Analysis of the velocity distribution function (VDF) and the energy distribution function (EEDF) of the electrons indicated the presence of highly energetic electrons near the cathode and a significant anisotropy of the VDF. 

Comparisons were carried out by executing simulations including and disregarding VUV photons. Remarkable differences for these different cases were found in the particle densities, fluxes, and discharge current pulse amplitudes. The analysis of photon trajectories indicated very strong trapping of the resonance radiation at the high pressure of 1 bar, as revealed by the spectral distribution of the arriving photons, as well. The strong trapping was found to result in a relatively low density (or number) of photons and this made the simulations feasible despite the extremely short time steps needed because of the high photoabsorption cross section.

The above, general findings regarding the importance of VUV photons in nanosecond, high-voltage discharges were found to hold at all parameter (pulse duration, voltage amplitude) values, which have been tested.
The variation of the gas pressure (that can change the transport characteristics of VUV radiation), the consideration of highly excited (Rydberg) states of helium (which were found to be present in significant concentration and to be responsible for a remarkable fraction of energy storage in the plasma \cite{Uwe}), as well as inclusion of photoionization of N$_2$ molecules by the VUV radiation can be identified as important future extensions of the present work.

\ack This work was supported by the National Office for Research, Development and Innovation (NKFIH) via grant 119357, Osaka University International Joint Research Promotion Program (Type A), the JSPS Grants-in-Aid for Scientific Research (S) 15H05736, and the UK EPSRC (EP/K018388/1) grant.

\end{document}